\documentclass[aps,
prb,
reprint,
floatfix,
superscriptaddress,
longbibliography,
nofootinbib,
]{revtex4-2}

\usepackage{graphicx}
\usepackage{physics}
\usepackage{booktabs}
\usepackage{amsthm,amssymb,mathrsfs}
\usepackage{tensor}
\usepackage{bm}
\usepackage[caption=false]{subfig}

\usepackage{feynmp-auto}
\usepackage{hyperref} 
\hypersetup{colorlinks=true,allcolors=blue}

\usepackage{xargs} 
\usepackage{xcolor}
\usepackage{slashed}

\usepackage[inline]{enumitem}

\usepackage{enumitem}
\setlist[itemize]{leftmargin=*}
\setlist[enumerate]{leftmargin=*}

\let\temp\phi
\let\phi\varphi
\let\varphi\temp

\newcommand{\iu}{\mathrm{i}} 
\newcommand{\e}{\mathrm{e}} 

\newcommand{\DD}{\mathscr{D}} 
\newcommand{\D}{\mathrm{d}} 
\newcommand{\sign}{\operatorname{sgn}} 
\providecommand{\ZZ}{\mathbb{Z}} 

\DeclareMathOperator{\arcosh}{arcosh}

\let\v\temp %
\let\temp\vv
\let\v\relax
\newcommand{\v}[1]{\ensuremath{\mathbf{#1}}} 

\makeatletter
\newsavebox{\@brx}
\newcommand{\llangle}[1][]{\savebox{\@brx}{\(\m@th{#1\langle}\)}%
	\mathopen{\copy\@brx\mkern2mu\kern-0.9\wd\@brx\usebox{\@brx}}}
\newcommand{\rrangle}[1][]{\savebox{\@brx}{\(\m@th{#1\rangle}\)}%
	\mathclose{\copy\@brx\mkern2mu\kern-0.9\wd\@brx\usebox{\@brx}}}
\makeatother

\makeatletter
\newcommand*{\coloneqq}{\mathrel{\rlap{%
			\raisebox{0.28ex}{$\m@th\cdot$}}%
		\raisebox{-0.28ex}{$\m@th\cdot$}}%
	=}
\newcommand*{\eqqcolon}{=\mathrel{\rlap{%
			\raisebox{0.28ex}{$\m@th\cdot$}}%
		\raisebox{-0.28ex}{$\m@th\cdot$}}%
}
\makeatother

\makeatletter
\newcommand{\dalembert}{\mathop{\mathpalette\dalembert@\relax}}
\newcommand{\dalembert@}[2]{%
	\begingroup
	\sbox\z@{$\m@th#1\square$}%
	\dimen0=\fontdimen8
	\ifx#1\displaystyle\textfont\else
	\ifx#1\textstyle\textfont\else
	\ifx#1\scriptstyle\scriptfont\else
	\scriptscriptfont\fi\fi\fi3
	\makebox[\wd\z@]{%
		\hbox to \ht\z@{%
			\vrule width \dimen0
			\kern-\dimen0
			\vbox to \ht\z@{
				\hrule height \dimen0 width \ht\z@
				\vss
				\hrule height 2\dimen0
			}%
			\kern-2\dimen0
			\vrule width 2\dimen0
		}%
	}%
	\endgroup
}
\makeatother

\makeatletter
\newcommand*{\rom}[1]{\expandafter\@slowromancap\romannumeral #1@}
\makeatother

\newcommand{\eqgraph}[3]{%
	\begin{gathered}
		\raisebox{0pt}[\dimexpr\height+#1][\dimexpr\depth+#2]{\ignorespaces#3\unskip}%
	\end{gathered}
}

\begin{document}
	
	\title{Quantum critical scaling of altermagnetism}
	
	\author{Sondre Duna Lundemo} 
	\affiliation{Center for Quantum Spintronics, Department of Physics, Norwegian University of Science and Technology, NO-7491 Trondheim, Norway}
	
	\author{Flavio S. Nogueira}
	\affiliation{Institute for Theoretical Solid State Physics, IFW Dresden, Helmholtzstr. 20, 01069 Dresden, Germany}
	
	\author{Asle Sudb\o}
	\email[Corresponding author: ]{asle.sudbo@ntnu.no}
	\affiliation{Center for Quantum Spintronics, Department of Physics, Norwegian University of Science and Technology, NO-7491 Trondheim, Norway}
	
	\date{\today} 
	
	\begin{abstract}
		The term altermagnetism has recently been introduced to describe the Néel order of a class of materials whose magnetic sublattices are neither related by translation nor inversion. 
		While these materials arguably have large technological potential, little effort has been devoted to studying the universal distinction of this phase of matter compared to collinear antiferromagnetism.
		Employing a recently proposed minimal microscopic model, we explicitly derive a nonlinear sigma model describing long-wavelength fluctuations of the staggered magnetization in this system, including quantum effects to leading order.
		The term that distinguishes the altermagnetic nonlinear sigma model from its antiferromagnetic counterpart is an interaction term that derives directly from the Berry phase of the microscopic spin degrees of freedom.
		Its effects on the one-loop renormalization group flow in $d=2+1$ dimensions are examined.
		Extending the theory to describe the fermionic excitations of the metallic altermagnet, we find an effective low-energy model of $d$-wave spin-split Dirac fermions interacting with the magnetic fluctuations. 
		Using a Dyson-Schwinger approach, we derive the many-body effects on the dynamical critical scaling due to the competition between the long-range Coulomb interaction and the fluctuations of the staggered magnetization. 
	\end{abstract}
	
	\maketitle 
	
	\section{Introduction}\label{sec:intro}

    The two-dimensional quantum antiferromagnet is one of the most studied systems in condensed matter physics \cite{chakravartyLowtemperatureBehaviorTwodimensional1988,chakravartyTwodimensionalQuantumHeisenberg1989,auerbachInteractingElectronsQuantum1994}. 
    The importance of this system is largely due to its profound connection to high-temperature superconductivity and the two-dimensional Hubbard model \cite{andersonResonatingValenceBond1987,affleckSU2GaugeSymmetry1988}.
	Not only has it fostered applications in quantum technologies and spintronics \cite{baltzAntiferromagneticSpintronics2018}, but also pathways to realizing exotic phases of matter that challenge our basic notion of order \cite{sachdevEffectiveLatticeModels1990,mudrySeparationSpinCharge1994,wenQuantumOrdersSymmetric2002}. 
    Following the discovery of the Berezinskii-Kosterlitz-Thouless transition, which demonstrated that singular fluctuations can drive a classical continuous transition \cite{kosterlitzOrderingMetastabilityPhase1973,nelsonUniversalJumpSuperfluid1977}, similar ideas were applied to the quantum antiferromagnet in $2+1$ dimensions \cite{haldaneO3Nonlinear$ensuremathsigma$1988,readValencebondSpinPeierlsGround1989}.
    The possibility of a continuous quantum phase transition between ground states of this system spontaneously breaking distinct global symmetries makes it ideal for confronting the Landau-Ginzburg-Wilson paradigm of phase transitions \cite{senthilDeconfinedQuantumCritical2004,senthilQuantumCriticalityLandauGinzburgWilson2004,senthilDeconfinedCriticalityCritically2005,kragsetFirstOrderPhaseTransition2006,kuklovDeconfinedCriticalityRunaway2006,kuklovDeconfinedCriticalityGeneric2008,herlandPhaseStructurePhase2013}.
    
	The discovery of the class of antiferromagnetic materials known as ``altermagnets" (AMs) has recently attracted significant interest due to their promising potential in technological applications \cite{smejkalEmergingResearchLandscape2022,smejkalConventionalFerromagnetismAntiferromagnetism2022,brekkeTwodimensionalAltermagnetsSuperconductivity2023,maeland_2024,roigMinimalModelsAltermagnetism2024}.
	Despite having no net magnetization, these materials display properties typically associated with ferromagnets. 
    Notably, they have spin-split electron bands without spin-orbit coupling \cite{leeBrokenKramersDegeneracy2024,krempaskyAltermagneticLiftingKramers2024,reimersDirectObservationAltermagnetic2024}. 
    These systems have so far mostly been analyzed through effective noninteracting fermion models. 
    While this provides insight into some properties of these materials, it fails to capture the inherently correlated nature of magnetism \cite{hertzQuantumCriticalPhenomena1976}.    
	An important problem left to address is how the altermagnetic properties arise from strong correlations and how they affect the fundamental aspects of the phase structure of the two-dimensional quantum antiferromagnet \cite{durrnagelAltermagneticPhaseTransition2024,heAltermagnetism$t$$t^prime$$delta$FermiHubbard2025,chenSpinExcitationsShastrySutherland2024,ferrariAltermagnetismShastrySutherlandLattice2024,giuliAltermagnetismInteractiondrivenItinerant2025,reDiracPointsTopological2024,kaushalAltermagnetismModifiedLieb2024}.
	The goal of this paper is to contribute to the latter.
	
	By employing a microscopic model of altermagnetism introduced in Ref.~\cite{brekkeTwodimensionalAltermagnetsSuperconductivity2023} we derive a long-wavelength effective theory of the fluctuations of the sublattice magnetization $\v{n}$ in this system, given by the following Euclidean Lagrangian (see Sec.~\ref{sec:derivation} for details)
	\begin{equation}\label{eq:ELagrangian_AM}
		\mathcal{L} = \frac{1}{2g}  (\partial_{\mu} \v{n})^2 +  \iu \theta \epsilon^{abc} n^{a} \partial_{\tau} n^{b} \partial_{x} \partial_{y} n^{c},
	\end{equation}
    subjected to the constraint $\v{n}^2 = 1$.
    Formulating this theory in the language of quantum field theory grants access to tools ideally suited to study phenomena that arise due to strong correlations. 
	In particular, we perform a renormalization group (RG) analysis of this theory in $d=2+1$ dimensions to elucidate the difference between the altermagnetic and antiferromagnetic phase transition. 
	By including fermions with hopping parameters consistent with the symmetries of the lattice, we expose a low-energy theory of spin-split Dirac cones. 
    With this theory, we compute the leading-order quantum fluctuation effects on the dynamical critical exponent $z$.
    
	\section{Derivation of long-wavelength field theory}\label{sec:derivation}
	
\begin{figure}[htb]
		\centering
		\includegraphics[width=0.9\columnwidth]{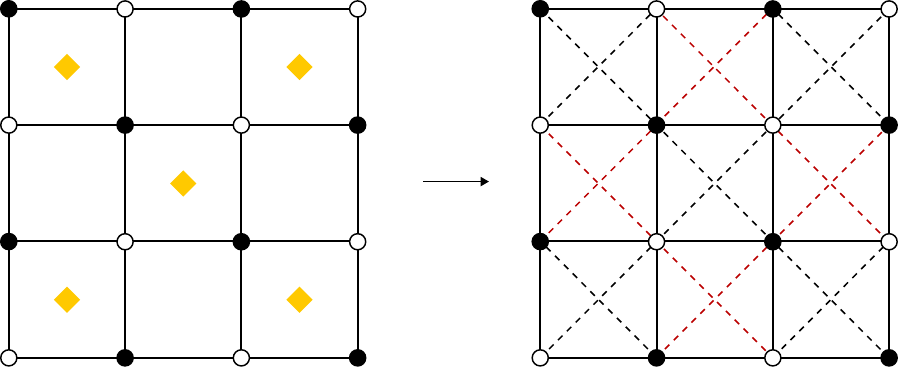}
		\caption{The Lieb lattice of the microscopic model, with the two magnetic sublattices indicated with filled and hollow dots respectively, and the nonmagnetic site as a yellow square. The nonmagnetic sites give rise to next-nearest-neighbor exchange couplings. These are illustrated by the dashed lines on the right-hand side: the red ones correspond to $-J'(1 - \gamma)$ while the black ones correspond to $-J'(1 + \gamma)$.}
		\label{fig:lieb_lattice}
	\end{figure}

	We describe the spin fluctuations of the altermagnet with the minimal model introduced in Refs.~\cite{brekkeTwodimensionalAltermagnetsSuperconductivity2023,maeland_2024}.
	The microscopic model constitute itinerant fermions on the Lieb lattice shown on the left-hand side of Fig.~\ref{fig:lieb_lattice}, where the center site (yellow squares) is a nonmagnetic site while the two other sites in the unit cell ($\bullet$ and $\circ$) carry a separate, localized spin degree of freedom.
	Without the itinerant fermions, the nonmagnetic site mediates via superexchange next-nearest-neighbor exchange couplings that are different on the two diagonals spanned by $\hat{\v{u}} \coloneqq 2^{-1/2} (\hat{\v{x}} + \hat{\v{y}}) $ and $\hat{\v{v}} \coloneqq 2^{-1/2} (\hat{\v{x}} - \hat{\v{y}})$.
	Combined with antiferromagnetic exchange couplings on nearest-neighbor links, the asymmetry induced by the nonmagnetic site changes the classical antiferromagnetic ground state of the square-lattice Heisenberg model into a so-called altermagnetic one. 
	The effective Hamiltonian of this spin system is given by 
	\begin{subequations}
		\begin{align}
			H &= \sum_{ij \in \Lambda} J_{ij} \v{S}_{i} \cdot \v{S}_{j},
		\intertext{where $\Lambda$ denotes the square lattice with lattice constant $a$, and the exchange couplings are}
			J_{i, i \pm \v{x}} &= J_{i, i \pm \v{y}} = J, \\
			J_{i,i\pm\bm{\delta}} &= - J' \Big[1 + \sign(\hat{\bm{\delta}})(-1)^{i}\gamma\Big], \quad \hat{\bm{\delta}} = \hat{\v{u}}, \hat{\v{v}}, \label{eq:exchange_nnn}
		\end{align}
	\end{subequations}
	where $\sign{\hat{\v{u}}} \coloneqq + 1$ and $\sign{\hat{\v{v}}} \coloneqq - 1$ , and $J, J' > 0$.
	The exchange couplings are illustrated graphically in Fig.~\ref{fig:lieb_lattice}.
	Following Ref.~\cite{brekkeTwodimensionalAltermagnetsSuperconductivity2023}, this model has been widely employed to describe altermagnetism \cite{yershovFluctuationinducedPiezomagnetismLocal2024,gomonayStructureControlDynamics2024,consoliSUNAltermagnetismLattice2025}, and is often referred to as a ``checkerboard model" \cite{canalsSquareLatticeCheckerboard2002}.   

	The long-wavelength effective action of the quantum antiferromagnet is obtained by performing a gradient expansion of the spin coherent state action appropriate for smooth fluctuations on top of an antiferromagnetic ground state, parametrized by \cite{affleckQuantumSpinChains1989,abanovTopologyGeometryQuantum2017}
	\begin{equation}\label{eq:spin_param}
		\langle \v{S}_{j} \rangle = S \left( (-1)^{j} \v{n}(x) + a \v{l}(x) \right),
	\end{equation} 
	where $x \coloneqq (\v{x},\tau)$ is the continuum coordinate corresponding to lattice site $j$ and Euclidean time $\tau$.
	The fields $\v{n}$ and $\v{l}$ are both assumed to be slowly varying fields on the lattice scale, and represent antiferromagnetic and ferromagnetic fluctuations respectively.
	As $\langle \v{S}_{j} \rangle^2 \simeq S^2$ we find to leading order in $a$ the local constraints $\v{n}^2 = 1$ and $\v{n} \cdot \v{l} = 0$.
	This yields a partition function given by the constrained functional integral
	\begin{subequations}\label{eq:partitionfunction}
		\begin{align}
			Z &\coloneqq \int \DD \v{n} \int \DD \v{l} \prod_{j} \delta( \v{n}^2_{j} - 1 ) \delta( \v{n}_{j} \cdot \v{l}_{j} )  \e^{-S} \\
			S[\v{n},\v{l}] &\coloneqq - 4\pi \iu S  \sum_{j} W_{0}[\langle \v{S}_{j} \rangle]+ \int \D\tau H(\v{n},\v{l}),
		\end{align}
	\end{subequations}
	where the first term in the Euclidean action represents the Berry phase of the quantum spin, with $ -4\pi W_{0}[\v{m}]$ being the solid angle encompassed by the classical field $\v{m}_j(\tau)$ during a period $\beta$ \cite{abanovTopologyGeometryQuantum2017}.
	In the continuum limit, we use $\sum_{j} \to a^{-D} \int \D^D x$ to obtain ($D\neq 1$)
	\begin{equation}
		\sum_{j} W_{0}[\langle \v{S}_{j}\rangle] \to \frac{D}{4\pi} \int \frac{\D^D x}{a^{D}} \int \D \tau a \v{l} \cdot \left(\v{n} \times \partial_{\tau} \v{n}\right).
	\end{equation}
	On nearest-neighbor links $\langle i, j \rangle $, using the prescription in Eq.~\eqref{eq:spin_param} with $\v{S}_{i} \cdot \v{S}_{i+\bm{\delta}} = (\v{S}_{i} + \v{S}_{i+\bm{\delta}})^2/2 - \v{S}_{i}^2/2 -  \v{S}_{i+\bm{\delta}}^2/2$ yields
	\begin{equation}
		J \sum_{\langle i,j \rangle} \v{S}_{i} \cdot \v{S}_{j} \to J \frac{S^2 a^2}{2} \int \frac{\D^2 x}{a^{2}}  \sum_{i=1,2} \Big[ \left(\partial_{i} \v{n}\right)^2 + 4 \v{l}^2 \Big].
	\end{equation}
	Next-nearest-neighbor links $\llangle i,j \rrangle$ couple moments on the same magnetic sublattice, which makes it more appropriate to use $\v{S}_{i} \cdot \v{S}_{i+\bm{\delta}} = -(\v{S}_{i} - \v{S}_{i+\bm{\delta}})^2/2 + \v{S}_{i}^2/2 +  \v{S}_{i+\bm{\delta}}^2/2$. 
	This yields
	\begin{widetext}
	\begin{equation}
			\sum_{\scalebox{0.7}{$\llangle$} i,j \scalebox{0.7}{$\rrangle$}} J_{ij} \, \v{S}_{i} \cdot \v{S}_{j} \to J'S^2 a^2 \int \frac{\D^2 x}{a^{2}} \left( \left(\left[\hat{\v{u}}\cdot\grad\right] \v{n}\right)^2 + \left(\left[\hat{\v{v}}\cdot\grad\right] \v{n}\right)^2\right)+ 2 J' \gamma S^2 a^3 \int \frac{\D^2 x}{a^{2}}  \, \v{l} \cdot \left( \left[\hat{\v{u}}\cdot\grad\right]^2 - \left[\hat{\v{v}}\cdot\grad\right]^2 \right) \v{n},
	\end{equation}
	where we have truncated the expansion at second order in spatial derivatives and neglected constant terms.
	Note that the spatially staggered part of the exchange coupling on the next-nearest-neighbor links in Eq.~\eqref{eq:exchange_nnn} is what makes the term linear in $\v{l}$ survive the continuum limit. 
	Thanks to
	\begin{equation}
		\left[\hat{\v{u}}\cdot\grad\right]^2 - \left[\hat{\v{v}}\cdot\grad\right]^2 = 2 \partial_{x} \partial_{y} \quad \text{and} \quad \left[\hat{\v{u}}\cdot\grad\right]^2 + \left[\hat{\v{v}}\cdot\grad\right]^2 = \partial_x^2 + \partial_y^2,
	\end{equation}
	and including the Berry phase, we conclude that
	\begin{equation}
		\begin{split}
			S[\v{n},\v{l}] \simeq \int \D^2 x \int \D \tau \left[ \frac{J S^2}{2} \left( 1 + 2 \frac{J'}{J} \right) \sum_{i=1,2} (\partial_i \v{n})^2 + 4 J S^2 \v{l}^2 + \v{l} \cdot \left( 4 J' \gamma S^2 a \partial_{x} \partial_{y} \v{n} - \iu \frac{2S}{a} \v{n} \times \partial_{\tau} \v{n} \right) \right]. 
		\end{split}
	\end{equation} 
	After including a Lagrange multiplier field $\lambda$ that enforces the local constraint $\v{n}\cdot \v{l} = 0$, the integral over $\v{l}$ can be performed straightforwardly.
	By demanding that $\delta S /\delta \lambda = 0$, we are left with
	\begin{equation}
		\begin{split}
			S_{\mathrm{eff}}[\v{n}] = \int \D^2 x \int \D \tau \bigg\{ \frac{1}{2g} \bigg[ (\partial_i \v{n})^2 + \frac{1}{c^2} (\partial_{\tau} \v{n})^2 \bigg] + J' \alpha \gamma^2 S^2 a^2 \left[ \left(\v{n} \cdot \partial_{x} \partial_{y} \v{n}\right)^2 - \left(\partial_{x} \partial_{y} \v{n}\right)^2 \right] + \iu \theta \v{n} \cdot \partial_{\tau} \v{n} \times \partial_{x}\partial_{y} \v{n} \bigg\}, 
		\end{split}
	\end{equation} 
	where summation over repeated indices is implied, and we have introduced $\alpha \coloneqq J'/J, g^{-1} \coloneqq J S^2 (1 + 2 \alpha), c^2 \coloneqq 2 J^2 S^2 a^2 (1 + 2 \alpha)$ and $\theta \coloneqq S \alpha \gamma $.
    Note that the novel term $\propto \theta$ derives directly from the cross term of the Berry phase of the spin and the spatially staggered part of the next-nearest-neighbor exchange after integrating out $\v{l}$.
	\end{widetext} 
	Keeping terms only to leading order in $\gamma$ yields the nonlinear sigma model (NLSM) for altermagnetism
	\begin{subequations}\label{eq:Eaction_AM}
		\begin{equation}
			S_{\mathrm{am}}[\v{n}] = \frac{1}{2g} \int \D^2 x \int \D \tau \left( \partial_{\mu} \v{n}\right)^2 + \iu \theta W[\v{n}],
		\end{equation}
		where $W[\v{n}]$ is the functional
		\begin{equation}\label{eq:W_functional}
			W[\v{n}] = \int\D^2 x \int \D \tau \epsilon^{abc} n^{a} \partial_{\tau} n^{b} \partial_{x} \partial_{y} n^{c},
		\end{equation}
	\end{subequations}
	and $\partial_{\mu} \coloneqq ( c^{-1} \partial_{\tau}, \bm{\nabla})$.
    The appearance of the mixed spatial derivatives in $W[\v{n}]$ reflects the description of altermagnetism as the simultaneous ordering of the Néel vector and some higher multipole moment of the magnetization \cite{bhowalFerroicallyOrderedMagnetic2024,mcclartyLandauTheoryAltermagnetism2024,mcclartyObservingAltermagnetismUsing2025}.
	Note that the functional $W[\v{n}]$ bears similarities with the Wess-Zumino-Witten (WZW) term in $1+1$ dimensions \cite{wittenNonabelianBosonizationTwo1984,affleckQuantumSpinChains1989}. 
	In fact, it exactly appears as a boundary term by performing partial integration with respect to one of the spatial coordinates in Eq.~\eqref{eq:W_functional}.
	However, $W[\v{n}]$ in total is not topological.
	
	\section{The nonlinear sigma model for altermagnetism}
	
	The term distinguishing the altermagnetic NLSM from its antiferromagnetic counterpart derived in the previous section was recently written down on phenomenological grounds \cite{gomonayStructureControlDynamics2024}.
	For completeness, let us derive the equations of motion and show that the model exhibits the $d$-wave splitting of spin-wave dispersions \cite{gomonayStructureControlDynamics2024}.
	For this purpose, it will be convenient to analytically continue the action to real time. 
	This yields
	\begin{subequations}
		\begin{align}
			- S_{\mathrm{am}}[\v{n}] &\to \iu \mathcal{S}_{\mathrm{am}}[\v{n}] = \iu \int \D^3 x \, \mathcal{L}_{\mathrm{am}}(\v{n}, \partial \v{n}), \label{eq:AMaction}
			\intertext{with}
			\begin{split}
				\mathcal{L}_{\mathrm{am}} &= \frac{1}{2g} (\partial_{\mu} \v{n}) \cdot \left( \partial^{\mu} \v{n}\right)	\\
				&\qquad\qquad+ \theta \epsilon^{abc} n^{a} \partial_{t} n^{b} \partial_{x} \partial_{y} n^{c},
			\end{split}
		\end{align}
		where we use the metric $\mathfrak{g} = \mathrm{diag}(1,-1,-1)$ and $\partial_{\mu} \coloneqq (c^{-1} \partial_{t}, \bm{\nabla})$.
	\end{subequations}
	The principle of least action $\delta \mathcal{S}_{\mathrm{am}}[\v{n}] = 0$ yields the Euler-Lagrange equations
	\begin{align}
			&\dalembert n^{a} = 2\theta g \epsilon^{abc} n^{b} \partial_{t} \partial_{x} \partial_{y} n^{c} \\
			&+ \theta g \epsilon^{abc} \left( \partial_{t} n^{b} \partial_{x} \partial_{y} n^{c} + \partial_{x} n^{b} \partial_{y} \partial_{t} n^{c} + \partial_{y} n^{b} \partial_{t} \partial_{x} n^{c} \right), \notag 
	\end{align} 
	with $\dalembert \coloneqq \partial_{\mu} \partial^{\mu}$.
	Linearizing these equations around a fixed ground state $\v{n}(x) = \v{n}_{0} + \delta \v{n}(x)$ yields 
	\begin{equation}
		\dalembert (\delta n^{a}) = 2 \theta g \epsilon^{abc} n^{b}_{0} \partial_{t} \partial_{x} \partial_{y} (\delta n^{c}).
	\end{equation} 
	Assuming a spin-wave form of the solutions $\delta n_{a} \sim \exp( \iu \omega t - \iu \v{k} \cdot \v{x} )$ yields the dispersion relation 
	\begin{equation}\label{eq:SW_dispersion}
		\omega_{\pm}(\v{k}) = \sqrt{c^2 \v{k}^2 + \left( \theta g c^2 k_x k_y \right)^2} \pm \theta g c^2 k_x k_y.
	\end{equation}
	The nondegeneracy of the spin-wave dispersions
	\begin{equation}\label{eq:magnon_splitting}
		\omega_{+}(\v{k}) - \omega_{-}(\v{k}) = 2 \theta g c^2 k_x k_y,
	\end{equation} 
	is directly related to the parameter $\theta$, which in turn is determined by the microscopic parameter $\gamma$ effecting the presence of the nonmagnetic site of the original model.
    The derivation in Sec.~\ref{sec:derivation} combined with Eq.~\eqref{eq:magnon_splitting} elucidates the quantum origin of the nondegenerate spin-wave spectrum. 
	Note that the $d$-wave momentum structure of the splitting is rotated $45^{\circ}$ with respect to that of the magnon bands in Ref.~\cite{brekkeTwodimensionalAltermagnetsSuperconductivity2023}.
    This amounts to a global rotation of the coordinate system used.
	
	\subsection{Renormalization group analysis in $2+1$ dimensions}\label{sec:RG}
	
	We are interested in the critical properties of the theory defined by the Euclidean action in  Eq.~\eqref{eq:Eaction_AM}.
	A first natural question to address is how the criticality of the ($2+1$)-dimensional NLSM is affected by $\theta \neq 0$.
	To this end, we follow closely the classical treatment of the NLSM in Refs.~\cite{chakravartyTwodimensionalQuantumHeisenberg1989,chakravartyTwodimensionalQuantumHeisenberg1989,zinn-justinQuantumFieldTheory2021}.  
	
	Consider the Euclidean Lagrangian in Eq.~\eqref{eq:ELagrangian_AM} $\mathcal{L} \coloneqq \mathcal{L}_{\mathrm{NLSM}} + \mathcal{L}_{W}$ and set $c=1$ in this section.
	To generate a perturbative expansion of the NLSM, we reparametrize the field $\v{n}$ in terms of a real scalar $\sigma(x)$ and the $(N-1)$-component field $\bm{\pi}(x)$ as $\v{n}(x) = \left( \sigma(x) \, \bm{\pi}(x) \right)^{\mathsf{T}}$, with $N=3$ corresponding to the physical case.
	After including a ``sublattice magnetic field" $\v{h}=(h \, \v{0})^{\mathsf{T}}$ for regularization purposes and resolving the local constraint exactly by using $\sigma = \sqrt{1 - \bm{\pi}^2}$, a standard manipulation yields \cite{zinn-justinQuantumFieldTheory2021}
	\begin{equation}\label{eq:Lafm}
		\begin{split}
			\mathcal{L}_{\mathrm{NLSM}}(\bm{\pi},\partial\bm{\pi})	&= \frac{1}{2g } \left( \partial_{\mu} \bm{\pi} \right)^2 + \frac{1}{2g} \frac{(\bm{\pi} \cdot \partial_{\mu} \bm{\pi} )^2}{1 - \bm{\pi}^2 } \\
			&- \frac{h}{g} \sqrt{1 - \bm{\pi}^2} + \frac{\rho}{2} \log\left( 1 - \bm{\pi}^2 \right), 
		\end{split}
	\end{equation}
    where $\rho$ denotes the number of degrees of freedom per unit volume \cite{nelsonMomentumshellRecursionRelations1977}.
	For small $g$, only fluctuations in $\bm{\pi}$ of order $\sqrt{g}$ are important \cite{zinn-justinQuantumFieldTheory2021}. 
	By rescaling $\bm{\pi} \mapsto \sqrt{g} \bm{\pi}$ and expanding to linear order in $g$, one gets
	\begin{equation}\label{eq:Lafm_scaled}
		\begin{split}
			\mathcal{L}_{\mathrm{NLSM}}(\sqrt{g}\bm{\pi},  \sqrt{g} \partial \bm{\pi}) &= \frac{1}{2} \left( \partial_{\mu} \bm{\pi} \right)^2 + \frac{g}{2} (\bm{\pi} \cdot \partial_{\mu} \bm{\pi} )^2 \\
			+ \frac{h}{2} \bm{\pi}^2 &+ \frac{hg}{8} \bm{\pi}^4  - \frac{\rho g}{2} \bm{\pi}^2 + \mathcal{O}(g^2).
		\end{split}
	\end{equation}
	Applying the same expansion to $\mathcal{L}_{W}$ with $N=3$ explicitly yields one new term to leading order in $g$
	\begin{align}\label{eq:LW}
			\mathcal{L}_{W}(\sqrt{g}\bm{\pi},  \sqrt{g} \partial \bm{\pi}) &= \iu g \theta \left( \partial_{\tau} \pi^{1} \partial_{x} \partial_{y} \pi^{2} - \partial_{\tau} \pi^{2} \partial_{x} \partial_{y} \pi^{1}  \right) \notag \\
			&+ \mathcal{O}(g^2).
	\end{align}
    Hence, after restoring the original $\bm{\pi}$ field, the quadratic part of the action in the Fourier basis can be written as 
    \begin{subequations}
        \begin{align}
            S_{\mathrm{am}}[\bm{\pi}] &= \frac{1}{2} \int \frac{\D^3 k }{(2\pi)^3} \pi_{i}(k)D^{-1}_{ij}(k) \pi_{j}(-k) + \dots,
            \intertext{where the propagator reads}
            \begin{split}\label{eq:Dpropagator}
                D(k) &= \frac{g}{(k^2 + h)^2 + (2 \theta g k_0 k_1 k_2 )^2} \\
            &\qquad\qquad \times \begin{pmatrix}
                k^2 + h & - 2 \theta g k_0 k_1 k_2 \\
                2 \theta g k_0 k_1 k_2 & k^2 + h
            \end{pmatrix}.
            \end{split}
        \end{align}
    \end{subequations}
    At this stage, the following can be asserted. 
    Since the one-loop diagrams contain one internal $D$ propagator, the $\theta$-dependence naturally comes with one extra power of $g$ compared to the well-known results with $\theta = 0$.
    To leading order in the perturbative expansion in powers of $g$, $\theta$ does not alter the one-loop RG flow of the NLSM.
    Nevertheless, we proceed with deriving the one-loop RG flow of the model for completeness.

    \begin{fmffile}{nlsm}
    	\fmfset{dash_len}{1.1mm}
    	\fmfset{wiggly_len}{2mm}
    	\fmfset{curly_len}{1mm}
    	\fmfset{dot_size}{3pt}
    	\fmfset{arrow_size}{2pt}

        We perform a one-loop RG analysis of the model directly in $d=2+1$ dimensions and at $T=0$ to investigate the effects of the new parameter $\theta$ on the Néel quantum critical point.
        Following \cite{zinn-justinQuantumFieldTheory2021}, we write the interaction vertex to leading order in $g$ diagrammatically as 
    	\begin{equation}
    		\eqgraph{1.5mm}{0ex}{
    			\begin{fmfgraph*}(30,20)
    				\fmfleft{l1,l2}
    				\fmfright{r1,r2}
    				\fmf{wiggly}{l1,v1,l2}
    				\fmf{dashes}{v1,v2}
    				\fmf{wiggly}{r1,v2,r2}
    		\end{fmfgraph*}}
    		\sim V^{(4)}_{i_{1} i_{2} i_{3} i_{4} } = \frac{1}{8g } \delta_{i_{1}i_{2}} \delta_{i_{3} i_{4}} \left[ (p_1 + p_2)^2 + h \right],
    	\end{equation}
    	where the dashed lines are only used to indicate the group index flow, and the wiggly line represents the $\bm{\pi}$ propagator.
    	The one-loop diagrams we have to consider are those in Fig.~\ref{fig:diagrams_nlsm}.
        These give no contribution to the renormalization of $\theta$.
        However, $\theta$ flows due to the flow of the spin rescaling factor.
        Moreover, the cancellation of the mass term from diagram~\ref{diag:a} with the shell contribution of the exponentiated functional integral measure shown in the last term of Eq.~\eqref{eq:Lafm} is not exact when $\theta \neq 0$. 
        However, it yields additional terms in the beta functions only at $\mathcal{O}(g^3)$, rendering them insignificant for the one-loop approximation, which is correct up to $\mathcal{O}(g)$. 
        These contributions are therefore omitted in the following.
        Evaluating the integrals over all frequencies and in the spatial momentum shell $\Lambda/\ell < \abs{\v{k}} < \Lambda$, we obtain the one-loop beta functions (see Appendix.~\ref{app:RG} for details) 
        \begin{subequations}\label{eq:beta_functions}
            \begin{align}
            \frac{\partial \hat{g}}{\partial \log \ell} &= - \hat{g} + \frac{N-2}{4\pi} \hat{g}^2 \\
            \frac{\partial \hat{h}}{\partial \log \ell} &= 2 \hat{h} + \frac{N-3}{8\pi} \hat{h} \hat{g} \\
            \frac{\partial \theta }{\partial \log \ell} &= - \frac{N-1}{4\pi} \theta\hat{g},
        \end{align}
        \end{subequations}
        where we have introduced the dimensionless coupling constants $\hat{g}\coloneqq \Lambda g$ and $\hat{h} \coloneqq h / \Lambda^2$.
        As anticipated, the parameter $\theta$ is perturbatively irrelevant, and it is clear that the well-known beta functions of the quantum NLSM are recovered as $\theta$ flows to $0$.
        However, judging from its connection with the WZW term alluded to above, there might be effects of the $\theta$ term beyond the perturbative RG \cite{wittenNonabelianBosonizationTwo1984}.

    	\begin{figure}[t]
    		\centering
    		\captionsetup[subfloat]{labelfont=bf}
    		\subfloat[\label{diag:a}]{
    			\centering
    			\begin{fmfgraph*}(40,25)
    				\fmfkeep{wave}
    				\fmfbottom{i,r}
    				\fmf{wiggly}{i,v1}
    				\fmf{dashes,tension=0.5}{v1,v2}
    				\fmf{wiggly}{v2,r}
    				\fmffreeze
    				\fmf{wiggly,left=1}{v1,v2}
    			\end{fmfgraph*}
    		}
    		~
    		\subfloat[\label{diag:b}]{
    			\centering
    			\begin{fmfgraph*}(40,30)
    				\fmfkeep{tadbpole}
    				\fmftop{t}
    				\fmfbottom{i,r}
    				\fmf{wiggly}{i,v1,r}
    				\fmffreeze
    				\fmf{dashes}{v1,v2}
    				\fmf{wiggly,left=1,tension=0.5}{v2,t,v2}
    			\end{fmfgraph*}
    		}
    		\caption{One-particle irreducible diagrams contributing to the one-loop order renormalization of the NLSM.
            } 
    		\label{fig:diagrams_nlsm}
    	\end{figure}	
    \end{fmffile}

	\section{Itinerant altermagnet}

    From the preceding section, we can conclude that the new term of the NLSM has no perturbative effects on the criticality of the Néel critical point to leading order in $g$.    
    This could have been anticipated by the fact that it contains more gradients compared to the other terms of the action. 
    To further investigate whether the criticality of altermagnets differs significantly from the antiferromagnetic counterpart, we include the fermionic excitations into the model.
    After all, the low-energy quasiparticles near the Fermi surface often dictate the quantum critical behavior in low-dimensional systems, in a way that is not captured by a bosonic critical theory \cite{sachdevQuantumPhaseTransitions2011}.

    \subsection{Lattice fermion model}

    \begin{figure*}[t]
        \subfloat[Three-dimensional plot of dispersion for $\alpha = \uparrow$. \label{dispersion:a}]{
            \centering
            \includegraphics[width=\columnwidth]{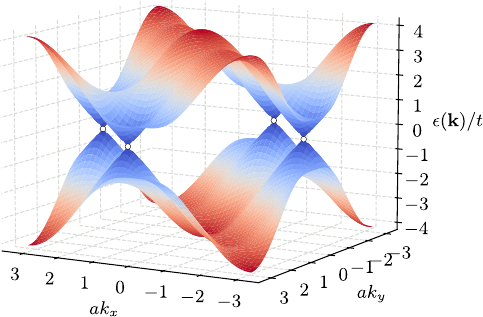}
        }
        \hfill
        \subfloat[Two-dimensional plot of dispersion for $\alpha=\uparrow$ and $\alpha = \downarrow$. \label{dispersion:b}]{
    		\centering
    		\includegraphics[width=\columnwidth]{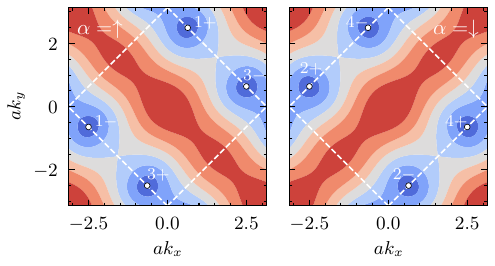}
        }
        \caption{Plot of dispersion in Eq.~\eqref{eq:fermion_dispersion} displaying the eight nodes. The parameters used in this plot are $t'= 0, \delta t = 0.7 t$, $\mu = 0$ and $J_{\mathrm{K}}S = t$. The dashed white lines on the right-hand panel show the lines in the BZ connecting the points $\mathrm{X}$ and $\mathrm{Y}$.}
		\label{fig:nodes}
	\end{figure*}

    Instead of re-introducing the nonmagnetic site and the hopping integrals in the model of Ref.~\cite{brekkeTwodimensionalAltermagnetsSuperconductivity2023}, we consider fermions hopping on the square lattice on the right-hand side of Fig.~\ref{fig:lieb_lattice} and choose the hopping parameters in a way that imprints the symmetries of the lattice. 
    In particular, we associate to each vertex $i \in \Lambda$ of the square lattice a fermionic degree of freedom with creation and annihilation operators $c^{\dagger}_{i\alpha}$ and $c^{\mathstrut}_{i\alpha}$ respectively.
    The fermions are governed by the Hamiltonian
    \begin{subequations}
        \begin{align}
            H_{\mathrm{f}} &= - \sum_{ij\in \Lambda}\sum_{\alpha} t_{ij}^{\mathstrut} c^{\dagger}_{i\alpha} c^{\mathstrut}_{j\alpha} - \mu \sum_{i \in \Lambda}\sum_{\alpha} c^{\dagger}_{i\alpha} c^{\mathstrut}_{i\alpha},
            \intertext{where $\mu$ denotes the chemical potential and $\alpha \in \{\uparrow,\downarrow\}$ is a spin label. The hopping parameters are}
            t_{i, i \pm \v{x}} &= t_{i, i \pm \v{y}} = t, \\
			t_{i,i\pm\bm{\delta}} &= t' + \sign(\hat{\bm{\delta}})(-1)^{i} \delta t , \quad \hat{\bm{\delta}} = \hat{\v{u}}, \hat{\v{v}}.\label{eq:hopping_nnn}
        \end{align}
    \end{subequations}
    The difference in hopping across the black and red diagonals of Fig.~\ref{fig:lieb_lattice} is supposed to mimic the quantum-chemical environment produced by the nonmagnetic site. 
    Variations of this type of model have been considered in Refs.~\cite{dasRealizingAltermagnetismFermiHubbard2024,boseAltermagnetismSuperconductivityMultiorbital2024,ferrariAltermagnetismShastrySutherlandLattice2024,giuliAltermagnetismInteractiondrivenItinerant2025,heAltermagnetism$t$$t^prime$$delta$FermiHubbard2025,consoliSUNAltermagnetismLattice2025,reDiracPointsTopological2024,meng-hanDiracPointsWeyl2024,antonenkoMirrorChernBands2025}.
    By introducing the Fourier-transform of the lattice fermions
    \begin{equation}
        c_{i\lambda \alpha} = \frac{1}{\sqrt{N}} \sum_{\v{k}\in \mathsf{BZ}} \e^{\iu\v{k}\cdot\v{r}_{i}} c_{\v{k} \lambda \alpha},
    \end{equation}
    and the two-component spinor $c_{\v{k}\alpha} \coloneqq \left( c_{\v{k} \bullet \alpha} \quad c_{\v{k} \circ \alpha } \right)^{\mathsf{T}}$ where $\lambda \in \{\bullet, \circ\}$ is a sublattice label, we can write the Hamiltonian as 
    \begin{equation}
	H_{\mathrm{f}} = \sum_{\v{k}\alpha} c_{\v{k}\alpha}^{\dagger} \mathbb{H}_{\v{k}\alpha}^{\mathstrut} c_{\v{k}\alpha}^{\mathstrut},
    \end{equation}
    where 
    \begin{subequations}
    \begin{align}
    	\mathbb{H}_{\v{k}\alpha} &= \begin{pmatrix}
    		- \Gamma_{\bullet}(\v{k}) - \mu & -\gamma(\v{k}) \\
    		-\gamma(\v{k}) & - \Gamma_{\circ}(\v{k}) - \mu
     	\end{pmatrix},
        \intertext{and}
    	\Gamma_{\lambda}(\v{k}) &\coloneqq \sum_{\hat{\bm{\delta}} \in \{ \pm \hat{\v{u}}, \pm \hat{\v{v}} \}} \left( t' + \sign{(\hat{\bm{\delta}})} (-1)^{\lambda} \delta t   \right) \e^{\iu \v{k} \cdot \bm{\delta}} \\
    	\gamma(\v{k}) &\coloneqq t \sum_{\hat{\bm{\delta}} \in \{ \pm \hat{\v{x}} , \pm\hat{\v{y}}\}} \e^{\iu\v{k}\cdot\bm{\delta}}
    \end{align}
    \end{subequations}
    At this point, we see that $\mathbb{H}_{\v{k} \alpha}$ is independent of $\alpha$, rendering the bands spin-degenerate and $\mathrm{C}_4$-invariant due to the equivalence of $\bullet$ and $\circ$ lattice sites. 
    This degeneracy is lifted by introducing Kondo coupling between the lattice fermions and the sublattice magnetization, which takes the form
    \begin{equation}\label{eq:HKondo}
	   H_{\mathrm{K}} = J_{\mathrm{K}} S \sum_{i \in \Lambda } \sum_{\alpha\beta} \sum_{\lambda, \lambda' \in \{ \bullet, \circ \}} c_{i\lambda\alpha}^{\dagger} \bm{\sigma}_{\alpha\beta} \tau^{3}_{\lambda \lambda'} c_{i\lambda' \beta}^{\mathstrut} \cdot \v{n}_{i},  
    \end{equation}
    where $\tau^{3}$ is the Pauli $z$ matrix acting on the sublattice indices, ensuring that $\v{n}$ is a smooth fluctuation on top of the staggered ground state. 
    Its mean-field part can be included in $H_{\mathrm{f}}$ to yield the altermagnetic $d$-wave splitting.
    Assuming without loss of generality that it points along the $z$ direction, we find
    \begin{equation}
	\mathbb{H}_{\v{k}\alpha} = \begin{pmatrix}
		-\Gamma_{\bullet}(\v{k}) - \mu + J_{\mathrm{K}} S \alpha  & -\gamma(\v{k}) \\
		-\gamma(\v{k}) & - \Gamma_{\circ}(\v{k}) - \mu - J_{\mathrm{K}} S \alpha
	\end{pmatrix},
    \end{equation}
    where $\alpha = +1 \,\hat{=} \uparrow$ and $\alpha = -1 \,\hat{=} \downarrow$.
    The eigenvalues of this matrix yield the fermion bands, which are given by
    \begin{equation}\label{eq:fermion_dispersion}
    	\begin{split}
    		\xi_{\v{k}\alpha \pm } = &- \mu - \left[ \Gamma_{\circ}(\v{k}) +  \Gamma_{\bullet}(\v{k}) \right] \\
            &\pm \sqrt{ (\gamma(\v{k}))^2 + \left[ \Gamma_{\bullet}(\v{k}) -  \Gamma_{\circ}(\v{k}) + \alpha J_{\mathrm{K}} S  \right]^2  }.
    	\end{split}
    \end{equation}
    Hence, the bands are spin-split whenever $\Gamma_{\circ} \neq \Gamma_{\bullet}$, which is the case when $\delta t \neq 0$.
    In fact, from
    \begin{equation}
        \Gamma_{\bullet}(\v{k}) -  \Gamma_{\circ}(\v{k}) = 4 \delta t \sin(k_x a) \sin(k_y a ),
    \end{equation}
    it is clear that this splitting has a $d$-wave momentum structure.

    \subsection{Low-energy Dirac theory}

    For some special values of the microscopic parameters, the fermion bands have nodes that are connected by exactly the magnetic ordering vector $\v{Q} = (\pi \quad \pi)^{\mathsf{T}}$.
    In particular, for $t' = 0$ and $\delta t > J_{\mathrm{K}} S/4$, ensuring the checkerboard symmetry of the lattice in Fig.~\ref{fig:lieb_lattice} is pronounced, the nodes lie exactly at $\epsilon(\v{k}) = 0$.
    The presence of such nodes was recently noted in Refs.~\cite{reDiracPointsTopological2024,meng-hanDiracPointsWeyl2024,antonenkoMirrorChernBands2025,heAltermagnetism$t$$t^prime$$delta$FermiHubbard2025}.
    This is illustrated in Fig.~\ref{fig:nodes}.
    The quasirelativistic behavior of the low-energy excitations in this regime makes the model interesting in many respects. 
    First, from a technical point of view, it facilitates the use of field-theoretical techniques.
    Second, it implies that the dynamical critical exponent $z$ that describes the relative scaling of the length and timescales is $1$ at the mean-field level. 
    While this scaling is also typical of insulating antiferromagnets, interaction with other degrees of freedom can alter it \cite{hertzQuantumCriticalPhenomena1976}.
    Moreover, the vanishing density of states of the low-energy excitations suppresses the screening of the Coulomb interaction. 
    Hence, it is anticipated that quantum fluctuations are important for the fate of these excitations.  
    In the following, we study this regime in more detail and set $a = 1$ for simplicity.

    By setting $t' = 0$ and restricting to the line $k_y = \pi - k_x$, it can be shown that the nodes of the dispersion for $\alpha = \uparrow$ are at
    \begin{equation}
        \tilde{k}_{x} = \pm \frac{1}{2} \arccos\left(1 - \frac{J_{\mathrm{K}}S}{2\delta t} \right) + n \pi,
    \end{equation}
    with $n\in\ZZ$.
    Hence, there are two nodes on this line, which are denoted by $\widetilde{\v{k}}_{1+}$ and $\widetilde{\v{k}}_{3-}$ according to Fig.~\ref{fig:nodes}, and which are given by
    \begin{equation}
                \widetilde{\v{k}}_{1+} = \left( k_* \quad \pi - k_* \right)^{\mathsf{T}} \, \text{and} \,\, \widetilde{\v{k}}_{3-} = \left( \pi -k_* \quad  k_* \right)^{\mathsf{T}},
    \end{equation}
    where $k_* \coloneqq \arccos\left(1 - J_{\mathrm{K}}S /(2\delta t) \right)/2$.
    By inserting $k_i = \widetilde{k}_{i, s} + p_{i}$ into the Hamiltonian and expanding in small $p_{i}$ one obtains
    \begin{equation}
        \mathbb{H}_{\widetilde{\v{k}}_{s} + \v{p}, \uparrow} = \mp v_{v\uparrow} \tau^{3} \left(\hat{\v{v}} \cdot \v{p}\right) +  v_{u\uparrow} \tau^{1} \left(\hat{\v{u}} \cdot \v{p}\right),
    \end{equation}
    where $v_{v\uparrow} \equiv v_1 \coloneqq \sqrt{2 J_{\mathrm{K}} S (4 \delta t - J_{\mathrm{K}} S )}$ and $v_{u \uparrow} \equiv v_2 \coloneqq t \sqrt{2 J_{\mathrm{K}} S / \delta t }$ and the upper sign refers to $s = 1+$ and the lower sign to $s = 3-$.
    The $\mathrm{C}_2$ symmetry of the lattice dictates that the velocity at the two opposite nodes $3+$ and $1-$ in Fig.~\ref{dispersion:b} are $\v{v}_{3+} = - \v{v}_{1+}$ and $\v{v}_{1-} = - \v{v}_{3+}$.
    By enlarging the basis with a new $2\times 2$ matrix grading that collects the low-energy modes at $\widetilde{\v{k}}$ and $-\widetilde{\v{k}}$ we can write the low-energy Euclidean Lagrangian in Dirac form
    \begin{align}
        \mathcal{L}_{\uparrow} &= \sum_{s=1,3} \bar{\bm{\psi}}_{s} \left( \gamma^{0} \partial_{0} + v_{1,s} \gamma^{1}\partial_{v} + v_{2,s} \gamma^{2}\partial_{u}   \right) \bm{\psi}_{s}, 
        \intertext{where the four-component spinors are}
        \bm{\psi}_{1} &\coloneqq \left(  \psi_{\uparrow1+} \quad \psi_{\uparrow 3 +} \right)^{\mathsf{T}} \, \text{and} \,\, \bm{\psi}_{3} \coloneqq \left(  \psi_{\uparrow1-} \quad \psi_{\uparrow 3 -} \right)^{\mathsf{T}},
    \end{align}
    and $v_{1,1} = - v_1$, $v_{1,3} = - v_1$, $v_{2,1} = - v_2$ and $v_{2,3} = +v_2$.
    The $4 \times 4$ gamma matrices are $\gamma^{0} \coloneqq \rho^{0} \otimes \tau^{2}$, $\gamma^{1} \coloneqq \rho^{3} \otimes \tau^{1}$ and $\gamma^{2} \coloneqq \rho^{3} \otimes \tau^{3}$.
    As usual, $\bar{\bm{\psi}} \coloneqq \bm{\psi}^{\dagger} \gamma^{0}$.

    Applying the same procedure for $\alpha = \downarrow$ and the nodes on the line $k_{y} = \pi + k_x$ yields a similar structure, which ultimately permits writing the low-energy Lagrangian for the fermion sector as
    \begin{equation}
    	\mathcal{L}_{\mathrm{f}} = \sum_{s=1}^{4} \bar{\bm{\psi}}_{s} \left( \gamma^{0} \partial_{0} + \v{d}_{s}(-\iu\grad) \cdot \bm{\gamma} \right) \bm{\psi}_{s},
    \end{equation}
    where
    \begin{equation}\label{eq:velocities}
	\v{d}_s(-\iu\grad) \coloneqq \begin{cases}
		(-v_1 \partial_{v}, - v_2 \partial_{u} ) & s = 1\\
		(+v_1 \partial_{u}, + v_2 \partial_{v} ) & s = 2\\
		(-v_1 \partial_{v}, + v_2 \partial_{u} ) & s = 3\\
		(+v_1 \partial_{u}, - v_2 \partial_{v} ) & s = 4
	\end{cases},
    \end{equation}
    and
    \begin{equation}
    	\bm{\psi}_{2} \coloneqq \left(  \psi_{\downarrow2+} \quad \psi_{\downarrow 4 +} \right)^{\mathsf{T}} \text{and} \,\, \bm{\psi}_{4} \coloneqq \left(  \psi_{\downarrow2-} \quad \psi_{\downarrow 4 -} \right)^{\mathsf{T}}.
    \end{equation}
    Indeed, with this expression, it can be shown that time reversal combined with a $\mathrm{C}_4$ spatial rotation maps the spin $\uparrow$ sector onto the spin $\downarrow$ sector, as expected for the altermagnet.

    \subsection{Combined theory of spin fluctuations and fermions}

    The two low-energy theories describing the fluctuations in the altermagnet are coupled through the Kondo interaction in Eq.~\eqref{eq:HKondo}.\footnote{Note that there is also a term arising from the Kondo Hamiltonian which couples the spin of the electrons to the ferromagnetic fluctuations $\v{l}$ introduced in Eq.~\eqref{eq:spin_param}, which leads to couplings between the electron spin and $\partial_{x} \partial_{y}\v{n}$ and $\v{n} \times \partial_{\tau} \v{n}$ upon integrating out $\v{l}$. However, the low-energy limit of these terms gives rise to interactions that are bi-quadratic in $\bm{\psi}$ and $\bm{\pi}$, and which include additional derivatives compared to \eqref{eq:LK}. We neglect these terms in the following.}
    Since the Dirac nodes are pairwise connected by the magnetic ordering vector $\v{Q}$, the theory is similar in spirit to the ``hot-spot" theory of spin density wave order in the cuprates \cite{abanovQuantumcriticalTheorySpinfermion2003,metlitskiQuantumPhaseTransitions2010,leeRecentDevelopmentsNonFermi2018}.
    However, by transforming to the Fourier basis and expanding the fermion momenta around the nodes, one realizes that the interaction between $\v{n}$ and the bilinears of $\bm{\psi}_{s}$ is local in the $s$ index.
    Since each node carries fermions with a definite spin, the slowly varying part of the interaction operates between $n^{3} = \sigma \equiv \sqrt{1 - \bm{\pi}^2}$ and the fermion bilinear $ (-1)^{s}\bar{\bm{\psi}}_{s} (\rho^{0} \otimes \iu \tau^{1}) \bm{\psi}_s$.
    The appropriate low-energy interaction therefore reads 
    \begin{equation}\label{eq:LK}
    	\mathcal{L}_{\mathrm{K}} = \kappa \bm{\pi}^2 \sum_{s=1}^{4} (-1)^{s} \bar{\bm{\psi}}_{s} \left(\rho^{0} \otimes \iu\tau^{1} \right) \bm{\psi}_{s},
    \end{equation}
    where $\iu \tau^{1} = \tau^{2} \tau^{3}$ comes from the staggering factor combined with changing $\bm{\psi}^{\dagger}$ to $\bar{\bm{\psi}}$ and $\kappa \sim J_{\mathrm{K}}$.
    Keeping only this leading term of the interaction ultimately yields the effective theory $\mathcal{L}_{\mathrm{eff}} \coloneqq \mathcal{L}_{\mathrm{f}} + \mathcal{L}_{\mathrm{K}} + \mathcal{L}_{\mathrm{NLSM}} + \mathcal{L}_{W}$.
    
	\subsection{Dynamical critical scaling}
    
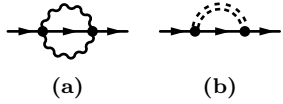
\begin{figure}[htb]
            \centering
            \captionsetup[subfloat]{labelfont=bf}
            \subfloat[\label{SDE:a}]{
                \centering
                \begin{fmfgraph*}(45,25)
        			\fmfkeep{Sigma1}
        			\fmfleft{i}
        			\fmfright{r}
        			\fmf{plain_arrow}{i,v1}
        			\fmf{plain_arrow,tension=0.7}{v1,v2}
        			\fmf{plain_arrow}{v2,r}
        			\fmffreeze
        			\fmf{boson,left=1}{v1,v2,v1}
        			\fmfdot{v1,v2}
        		\end{fmfgraph*}
            }
            ~
            \subfloat[\label{SDE:b}]{
                \centering
                \begin{fmfgraph*}(45,25)
        			\fmfkeep{Sigma2}
        			\fmfleft{i}
        			\fmfright{r}
        			\fmf{plain_arrow}{i,v1}
        			\fmf{plain_arrow,tension=0.7}{v1,v2}
        			\fmf{plain_arrow}{v2,r}
        			\fmffreeze
        			\fmf{dbl_dashes,left=1}{v1,v2}
        			\fmfdot{v1,v2}
        		\end{fmfgraph*}
            }
            \caption{Fermionic self energies due to the spin fluctuations \textbf{(a)} and the Coulomb interaction \textbf{(b)}. The plain line denotes fermion propagators, the wiggly line the $\bm{\pi}$ propagator and the double-dashed lines the dressed $\phi$ propagator.} 
            \label{fig:SDE_diagrams}
    	\end{figure}

    Since the leading interaction between $\bm{\pi}$ and $\bm{\psi}_s$ is biquadratic, there will be no new contributions to the one-loop RG flow of the NLSM.
    However, at two-loop order, there is a fermionic self-energy due to $\mathcal{L}_{\mathrm{K}}$ that possibly alters the dynamics of the fermions. 
    This is illustrated in Fig.~\ref{SDE:a}.
    The effects of this interaction can be studied using the Dyson-Schwinger (DS) equation for the dressed fermion propagator.
    In general, this can lead to spontaneous mass generation of the fermions, driving a metal-insulator transition in this case, or a modification of the quasirelativistic structure of the low-energy excitations, which is not protected by Lorentz invariance. 
    In the following, we focus on the latter and shed some light on the former in Appendix~\ref{app:mass_generation}.
    Based on the perturbative irrelevance of the interaction term in $\mathcal{L}_{W}$, we set $\theta = 0$ in this calculation.
    
    \begin{fmffile}{SDE}
    		\fmfset{dash_len}{1.5mm}
        	\fmfset{wiggly_len}{2mm}
        	\fmfset{curly_len}{1mm}
        	\fmfset{dot_size}{3pt}
        	\fmfset{arrow_len}{6pt}
    \end{fmffile}

    The vanishing density of states of the low-energy fermionic excitations implies that the long-range electromagnetic interaction mediated by a $\mathrm{U}(1)$ gauge field in $3+1$ dimensions is not screened  \cite{herbutQuantumCriticalPoints2001,herbutInteractionsPhaseTransitions2006,sonQuantumCriticalPoint2007,hsiaoDualityUniversalTransport2017}.
    In a nonrelativistic limit, the fluctuations of the spatial components of the gauge field are suppressed and the interaction is given by the instantaneous Coulomb interaction, mediated by the scalar potential $\phi$. 
    The coupling between the low-energy fermions and $\phi$ is of a Yukawa type and gives rise to the self-energy diagram in Fig.~\ref{SDE:b}. 
    Here, the double-dashed line denotes the propagator of the scalar potential dressed with the fermionic polarization bubble. 
    The vertex corrections in these diagrams are neglected.
    Specifically, the Euclidean Lagrangian we consider in the following is given by
    \begin{align}
    		&\mathcal{L}_{\mathrm{eff}} = \sum_{s=1}^{4} \bar{\bm{\psi}}_{s} \left( \gamma^{0} \partial_{0} + \v{d}_{s}(-\iu\grad) \cdot \bm{\gamma} + \iu e \phi \gamma^{0}  \right) \bm{\psi}_{s}  \\
    		&+\frac{1}{2g} \left( \partial_{\mu} \bm{\pi} \right)^2 +\kappa \bm{\pi}^2 \sum_{s=1}^{4} (-1)^{s} \bar{\bm{\psi}}_{s} \left(\rho^{0} \otimes \iu\tau^{1} \right) \bm{\psi}_{s} + \frac{1}{2} \phi \abs{\nabla} \phi, \notag
    \end{align}
    where $\phi$ denotes the electric scalar potential, and the notation $\abs{\nabla}$ is schematic for the kernel $\abs{\v{q}}$ in momentum space \cite{herbutQuantumCriticalPoints2001}.
    Here, we have explicitly disregarded the self-interactions of the $\bm{\pi}$ field of the NLSM. 

    By computing the diagrams in Fig.~\ref{fig:SDE_diagrams} explicitly in $d=3$ with $\abs{v_1} = \abs{v_2} = v$ and $\hat{v}$ denoting the fermion velocities in units of the $\bm{\pi}$ boson velocity $c$ (see Appendix~\ref{app:self_energy} for details) we arrive at the following equation for the renormalization of the velocities $v_{i}(q) = v_i (1 - \gamma_{i}(\hat{v},\lambda,u) \log \Lambda/q )$,
    where the anomalous dimensions $\gamma_{i}$ are given by
    \begin{subequations}\label{eq:anomalous_dims}
            \begin{align}
            \gamma_{1}(\hat{v},\lambda,u) &= u \hat{v}^2 f_{i}(\hat{v}) - \frac{2}{\pi^2 N_s} \Theta_{i}(\lambda)\\
            \gamma_{2}(\hat{v},\lambda,u) &= u (\hat{v}^2 f_{i}(\hat{v}) - f_{0}(\hat{v})) - \frac{2}{\pi^2 N_s} \Theta_{i}(\lambda).
            \end{align}
    \end{subequations}
    The dimensionless coupling constants appearing here are $u \coloneqq \kappa^2 g^2 (N-1)/(16\pi^2)$ and $\lambda \coloneqq e^2/(16 N_s v)$ and $N_s$ is the number of fermion flavors.
    The functions $f_{0}(\hat{v})$, $f_{i}(\hat{v})$ and $\Theta_{i}(\lambda)$ are defined in Eqs.~\eqref{eq:f0}, \eqref{eq:f1} and \eqref{eq:Theta} respectively in Appendix~\ref{app:self_energy}.

     \begin{figure}[htb]
    	\centering
    	\includegraphics[width=\columnwidth]{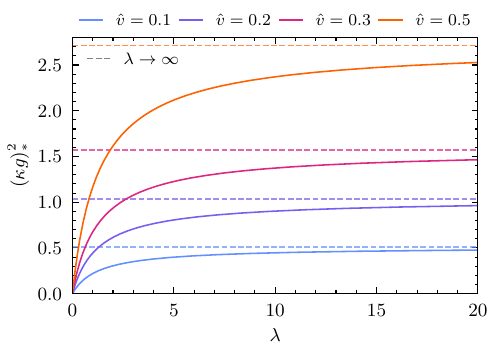}
    	\caption{Critical coupling in Eq.~\eqref{eq:critical_kappa} that makes $z = 1$ as a function of $\lambda$, for a selection of values of $\hat{v}$ and with $N_s = 4$ and $N = 3$. The dashed lines indicate the limit of infinite Coulomb interaction strength. }
    	\label{fig:critical_coupling}
    \end{figure} 
    
    At this stage, some remarks are in order. 
    Even though all the Dirac cones are isotropic at the outset, i.e., $\abs{v_1} = \abs{v_2} = v$, the algebraic structure of the interaction derived from the Kondo Hamiltonian leads to $\gamma_{1} \neq \gamma_2$.  
    Upon noticing that $\partial_{u}$ and $\partial_{v}$ swap roles for $\alpha = \uparrow$ and $\alpha = \downarrow$ in Eq.~\eqref{eq:velocities}, we conclude that quantum fluctuations generically drives $\abs{v_1} \neq \abs{v_2}$, creating an anisotropy consistent with the symmetries of the altermagnetic lattice model. 
    Specifically, under a rescaling of momenta $\v{p} \mapsto \ell \v{p}$, the dispersion of the fermions scale as
    \begin{equation}
		\omega \mapsto \omega(\ell) = \ell^{z} \sqrt{ (v p_1)^2 + \ell^{-2\Delta} (vp_2)^2 },
	\end{equation}
    with $z = 1 + \max\left\{\gamma_{1},\gamma_{2}\right\} = 1 + \gamma_{1}$, and $\Delta = \gamma_1 - \gamma_2$, and where $p_1$ and $p_2$ refer to the momenta corresponding to the first and second component of the vector $\v{d}_{s}$ in Eq.~\eqref{eq:velocities} respectively.
    The anisotropic scaling $\Delta \neq 0$ is insensitive to the Coulomb interaction and derives directly from the matrix $\rho^{0} \otimes \iu\tau^{1}$ appearing in the interaction vertex of $\bm{\pi}^2$ and the fermion bilinear.
    Moreover, we see that the magnetic fluctuations and the Coulomb interaction have competing effects on the dynamical critical exponent $z$.
    Hence, there is a critical value of the coupling constant $(\kappa g)^2$ that restores the mean-field result $z=1$.
    This critical value is given by
    \begin{equation}\label{eq:critical_kappa}
        \left( \kappa g \right)^2_{*}  = \frac{32}{N_s (N-1)} \frac{\Theta_i(\lambda)}{\hat{v}^2 f_{i}(\hat{v})},
    \end{equation}
    which is plotted as a function of $\lambda$ in Fig.~\ref{fig:critical_coupling}.
    Microscopically, this can be traced back to the ratio of the Kondo coupling and the nearest-neighbor exchange $\left( \kappa g \right)^2 \sim (J_{\mathrm{K}}/J)^2$.
    In particular, this critical value is consistent with working perturbatively in $\kappa g$ only if $\hat{v}$ is small and/or $\lambda$ is small. 
    Generally, however, we expect this competition to yield a deviation from $z=1$, albeit a small one.
    Already at the level of the spin-wave spectrum of the nonlinear sigma model in Eq.~\eqref{eq:SW_dispersion} it can be inferred that $\theta \in [0,\infty)$ interpolates between $z=1$ and $z=2$.
    Interestingly, even with $\theta = 0$ and including some pronounced features of altermagnetism in the low-energy fermion theory, we generically find $z > 1$ due to the fluctuations of the sublattice magnetization.
    However, from Fig.~\ref{fig:critical_coupling} and the discussion preceding it, one expects the contribution from the Coulomb interaction to dominate for correlated fermions, making $z < 1$ as in Ref.~\cite{sonQuantumCriticalPoint2007}.
    The limit $u \to 0$ in Eq.~\eqref{eq:anomalous_dims} restores the results of Ref.~\cite{sonQuantumCriticalPoint2007} for the low-energy excitations of graphene. However, one should bear in mind that this limit is somewhat artificial because the nodal structure of the low-energy fermions in our case owes its very existence to $J_{\mathrm{K}} \neq 0$.
	
	\section{Discussion and conclusion}\label{sec:Conclusion}

    To summarize, we have derived and studied a field theory for the long-wavelength fluctuations of an altermagnet, starting from a minimal microscopic model.
    While originally introduced as a toy model in Ref.~\cite{brekkeTwodimensionalAltermagnetsSuperconductivity2023}, recent experiments performed on compounds with the same Lieb-lattice structure have unravelled hallmark properties of altermagnetism \cite{zhangCrystalsymmetrypairedSpinvalleyLocking2024}.
    This suggests that the theory developed in this paper may prove relevant to describing real materials.
    We find that the term distinguishing the altermagnetic NLSM from the antiferromagnetic counterpart does not perturbatively modify the one-loop RG flow. 
    The high-loop order needed to see its effects suggests that the altermagnetic phase transition is likely not universally different from the Néel quantum critical point.
    Since the lattice symmetries of the altermagnet impose constraints on the allowed dimer states of the model, an interesting subject for future studies might be the nature of the Néel--valence-bond-solid transition in altermagnets. 
    An ideal starting point for such a study is the nonlinear sigma model we have derived.

    Since most proposed models of altermagnets involve itinerant fermions, we extended the theory with a fermion sector dictated by the lattice symmetries. 
    With a finite staggered mean-field magnetization, we revealed a low-energy theory consisting of four flavors of four-component Dirac spinors in a certain parameter regime. 
    By deriving the fermionic Dyson-Schwinger equations, taking into account both magnetic fluctuations and the Coulomb interaction, we found competing effects on the dynamical critical scaling. 
    In particular, while the Coulomb interaction tends to reduce $z$ from its mean-field value $z_{\mathrm{MFT}} = 1$, the magnetic fluctuations tend to increase it.
    The situation is remarkably similar to the scenario proposed in Ref.~\cite{herbutQuantumCriticalPoints2001}, where such a competition is conjectured to restore $z=1$ exactly, based on gauge invariance and the naive scaling of $\phi$. 
    However, since our results are obtained directly in $d=3$ and survive even as the charge $e^2 \to \infty$, the argument of Ref.~\cite{herbutQuantumCriticalPoints2001} does not apply.
    Generically, we expect interactions in this model to yield a dynamical critical exponent deviating slightly from $1$, depending on the details in the parameters of the model at the critical point.
    
    Apart from the nontrivial dynamical scaling, the spin fluctuations in this model induce an anisotropy in the scaling of momenta in different directions for each node in a way that pronounces the symmetries of the altermagnet. 
    Although originally arising from a lattice model with these symmetries imprinted in the coupling constants, it manifests itself only through an algebraic structure in the low-energy effective field theory. 

	\begin{acknowledgments}
        We thank U. Seifert for useful discussions.
		We acknowledge support from the Norwegian Research Council through Grant No. 262633, ``Center of Excellence on Quantum Spintronics” and Grant No. 323766, as well as COST Action CA21144  ``Superconducting Nanodevices and Quantum Materials for Coherent Manipulation".
	\end{acknowledgments}

    \appendix

    \section{One-loop RG of the nonlinear sigma model}\label{app:RG}
     
    In this Appendix, we provide some details on the RG analysis used to obtain the beta functions Eqs.~\eqref{eq:beta_functions}.
    Following Ref.~\cite{chakravartyTwodimensionalQuantumHeisenberg1989}, we use a momentum-shell method to derive the recursion relations for the parameters of the theory in fixed dimension $d=2+1$. 
    We perform the integral over all frequencies at $T=0$ and the spatial integrals in a momentum shell. 
    Within the accuracy of the one-loop expansion, the recursion relations read \cite{chakravartyTwodimensionalQuantumHeisenberg1989}
    \begin{equation}
    	\frac{1}{g'} = \zeta^2 \ell^{-d-2} \left[ \frac{1}{g} + I_{\mathrm{loop}} \right],
    \end{equation}
    and 
    \begin{equation}
    	\left( \frac{h}{g} \right)^{\prime} = \zeta^2 \ell^{-d} \left[ \frac{h}{g} + h \frac{N-1}{2} I_{\mathrm{loop}} \right],
    \end{equation}
    while $\theta' = \zeta^2 \ell^{-d-3} \theta$. 
    The spin rescaling factor $\zeta$ is determined by demanding that rotational invariance is preserved under rescaling  \cite{nelsonMomentumshellRecursionRelations1977}
    \begin{equation}
        \zeta = \ell^{d} \left[ 1- \frac{g}{2} \left(N-1\right) I_{\mathrm{loop}} \right].
    \end{equation}
    The loop integral appearing here differs from that of the conventional NLSM due to the parameter $\theta$ and reads
    \begin{equation}
    	I_{\mathrm{loop}} = \int \frac{\D^3 k}{(2\pi)^3} \frac{(k^2 + h)}{(k^2 + h)^2 + (2 \theta g)^2 (k_0 k_1 k_2)^2 }.
    \end{equation}
    The computation is facilitated by splitting the integral in the following fashion
    \begin{align}
    	I_{\mathrm{loop}} = \frac{1}{2} \int \frac{\D^3 k}{(2\pi)^3} \left[ \frac{1}{k^2 + h + 2 \iu\theta g k_0 k_1 k_2 } + \mathrm{c.c.} \right],
    \end{align}
    after which each term can be computed by using a Schwinger parametrization.
    Denoting $I_{\mathrm{loop}} = (\mathcal{I} + \bar{\mathcal{I}})/2$ one finds that
    \begin{align}
    	\mathcal{I} &= \int \frac{\D^2 k}{(2\pi)^2} \int \frac{\D \omega}{2\pi} \int_{0}^{\infty} \D \lambda \e^{-\lambda \left( \omega^2 + \v{k}^2 + h + 2\iu \theta g \omega k_1 k_2 \right)} \notag \\
    	&= \int \frac{\D^2 k}{(2\pi)^2} \int_{0}^{\infty} \frac{\D \lambda}{\sqrt{4\pi \lambda}} \e^{- \lambda (\v{k}^2 + h)} \e^{-\lambda (\theta g)^2 (k_1 k_2)^2}.
    \end{align}
    At this stage, both the angular integral and the momentum-shell integral can be performed. 
    After transforming to dimensionless integration variables $\lambda = \xi/\Lambda^2$, one obtains
    \begin{equation}
        I_{\mathrm{loop}} = \Lambda \log(\ell) \mathcal{J}(\theta \hat{g},\hat{h}),
    \end{equation}
    where
    \begin{equation}
    \begin{split}
        \mathcal{J}(\theta \hat{g},\hat{h}) = \frac{1}{\sqrt{2}(2\pi)^{3/2}} \int_{0}^{\infty} \frac{\D \xi}{\sqrt{\xi}} &\e^{-\xi(1+\hat{h}) - \xi (\theta \hat{g})^2/8 } \\
        &\times I_0\left( \frac{\xi(\theta \hat{g})^2}{8} \right),
    \end{split}
    \end{equation}
    where $I_{0}(x)$ denotes the modified Bessel function of the first kind of order zero.
    Although the function $\mathcal{J}(\theta \hat{g},\hat{h})$ contains a dependence on $\theta$, this comes with higher powers of $\hat{g}$ which puts these corrections beyond the level of accuracy of the one-loop calculation. 
    Therefore, we should evaluate $\mathcal{J}(\theta\hat{g},\hat{h})$ at $\hat{g}=0=\hat{h}$, yielding $\mathcal{J}(0,0) = 1/(4\pi)$, from which Eqs.~\eqref{eq:beta_functions} follow.

    \onecolumngrid
    \section{Details on self-energy calculations}\label{app:self_energy}
    In this Appendix, we provide details on the DS analysis used to arrive at Eqs.~\eqref{eq:anomalous_dims}.
    For completeness, we also include details on the computation of the self-energies from the Coulomb interaction, which is known in the literature \cite{sonQuantumCriticalPoint2007,bauerNonperturbativeRenormalizationGroup2015}.
    To compute the dressed propagator of the scalar potential, we first need the fermion polarization bubble, $\Pi(q)$.
    The dependence on the node $s$ enters only in the square of the velocities $v_{1}$ and $v_2$, which are assumed to be equal in magnitude for this calculation.
    To simplify further, we can set $v=1$ throughout the computation of $\Pi(q)$ and reinstate it at the end.
    The bubble reads
    
    \begin{align}
    	\Pi(q) &= + e^2\sum_{s} \int \frac{\D^3 k }{(2\pi)^3} \tr \frac{1}{\iu \slashed{k}} \iu \gamma^{0} \frac{1}{\iu(\slashed{k} - \slashed{q})} \iu \gamma^{0}.
    \end{align}
    Using $\tr (\gamma_{\mu} \gamma_{0} \gamma_{\nu} \gamma_{0}) = 4 \left( 2 \delta^{\mu 0} \delta^{\nu 0} - \delta^{\mu\nu} \right)$ we find
    \begin{align}
    	\Pi(q) &= 4 N_s e^2  \int_{0}^{1} \D x \int \frac{\D^3 k }{(2\pi)^3} \frac{2 k_0 (k_0 - q_0) - k \cdot (k-q)}{(\ell^2 + \Delta)^2},
    \end{align}
    where $N_s$ is the number of nodes, $\ell \coloneqq k - qx$ and $\Delta = q^2 x(1-x)$.
    
    By using the spherical symmetry of the remaining integral over $\ell$, we can remove all the linear terms of the numerator and perform the replacement $(\ell_{0})^2 \to \ell^2/d$ \cite{sonQuantumCriticalPoint2007}. 
    This yields
        \begin{align}
    	   \Pi(q) &= 4 N_s e^2  \int_{0}^{1} \D x \int \frac{\D^d \ell}{(2\pi)^d} \frac{(2-d) \ell^2/d - 2 q_{0}^2 x(1-x) + q^2 x(1-x)}{(\ell^2 + \Delta)^2} \notag \\
           &= \frac{4 N_s e^2}{(4\pi)^{d/2}}  \int_{0}^{1} \D x \bigg[ \Gamma(2 - d/2) \left[ q^2 - 2 q_{0}^2 \right] x(1-x) q^{d-4} (x(1-x))^{d/2-2} + \frac{2-d}{d} \frac{d}{2} \Gamma(1 - d/2) q^{d-2} (x(1-x))^{d/2-1}  \bigg] \notag \\
        	&= 8 N_s e^2 \frac{1}{(4\pi)^{d/2}} \Gamma(2 - d/2) \abs{\v{q}}^2 \abs{q}^{d-4} \int_{0}^{1} \D x (x(1-x))^{d/2-1} \notag \\
            &= \frac{e^2}{8 v^2} N_s \frac{v^2\v{q}^2}{\sqrt{q_{0}^2 + v^2 \v{q}^2}},
        \end{align}
    where we have let $d\to 3$ and restored the velocity $v$ in the last step.
    
    The dressed $\phi$ propagator therefore reads
    \begin{equation}\label{eq:rescaled_prop}
    	\mathcal{D}(q) =  \left[ 2 \abs{\v{q}} + \frac{e^2}{8} N_s \frac{\v{q}^2}{\sqrt{q_{0}^2 + v^2 \v{q}^2}} \right] ^{-1}.
    \end{equation}
    By introducing the dimensionless coupling constant $\lambda \equiv e^2/(16 N_s v)$ as in Ref.~\cite{sonQuantumCriticalPoint2007}, we can write the propagator evaluated at rescaled momenta $\v{q} \mapsto \v{q}/v$ as 
    \begin{align}\label{eq:Drescaled}
        \widetilde{e^2 \mathcal{D}(q)} = \frac{8 v^2}{N_s} \lambda \left[ \abs{\v{q}} + \lambda \frac{\v{q}^2}{\abs{q}} \right]^{-1}.
    \end{align}

    The anisotropic momentum structure induced by the diagrams in Fig.~\ref{fig:SDE_diagrams} requires that three different wave-function renormalizations $Z_{\mu}$ are introduced. 
    That is, we write the dressed fermion propagator as 
    \begin{equation}
        \mathscr{G}_{s}(q) = \frac{1}{Z_{0}(q) \gamma^{0} \iu q_{0} + Z_{1}(q) v_{1} \gamma^{1} \iu q_1 + Z_{2}(q) v_{2} \gamma^{2} \iu q_2 }.
    \end{equation}
    The DS equation without vertex corrections for the dressed fermion propagator now reads
    \begin{equation}
    	\mathscr{G}^{-1}_{s}(q) = 	\mathscr{G}^{-1}_{s 0}(q) + \Sigma_{\bm{\pi},s}(q) + \Sigma_{\phi,s}(q) \equiv \mathscr{G}^{-1}_{s 0}(q) + \Sigma(q),
    \end{equation}
    where $\Sigma_{\bm{\pi},s}(q)$ and $\Sigma_{\phi,s}(q)$ are the amputated self-energy diagrams of Fig.~\ref{fig:SDE_diagrams} but with the internal fermion propagators replaced by the dressed ones.
    After rescaling the spatial momenta by $1/c$, we find that
    \begin{equation}
    \begin{split}
        \Sigma_{\bm{\pi},s}(q) = -\frac{2  \kappa^2 (g c^2)^2 (N-1)}{c^4} \int \frac{\D^3 k}{(2\pi)^3} & \int \frac{\D^3 p}{(2\pi)^3} \frac{1}{k^2 (k + p- q)^2} \\
        &\times(\rho^{0} \otimes \iu \tau^{1}) \frac{1}{Z_{0}(p) \gamma^{0} \iu p_{0} + Z_{1}(p) \hat{v}_{1} \gamma^{1} \iu p_{1} + Z_{2}(p) \hat{v}_{2} \gamma^{2} \iu p_{2}  } (\rho^{0} \otimes \iu \tau^{1}),
    \end{split}
    \end{equation}
    where $\hat{v}_1$ and $\hat{v}_2$ are now the fermion velocities in units of the $\bm{\pi}$ boson velocity $c$. 
    From Eq.~\eqref{eq:velocities} it is clear that $p_1$ and $p_2$ in this equation correspond to either $\hat{\v{u}} \cdot \v{p}$ and $\hat{\v{v}} \cdot \v{p}$ depending on the node $s$ in question. 
    However, since the $\bm{\pi}$ boson propagator is isotropic, we can always rotate the coordinate system so that $p_1$ and $p_2$ correspond to $p_x$ and $p_y$, respectively.
    Moreover, the integral over $k$ can in this case be performed straightforwardly \cite{lundemoTopologydrivenDeconfinedQuantum2025}, and it yields
    \begin{equation}\label{eq:Sigma_pi}
        \Sigma_{\bm{\pi},s}(q) = - 4\pi^2 u \int \frac{\D^3 p}{(2\pi)^3} \frac{1}{\abs{p-q}} (\rho^{0} \otimes \iu \tau^{1}) \frac{1}{Z_{0}(p) \gamma^{0} \iu p_{0} + Z_{1}(p) \hat{v}_{1} \gamma^{1} \iu p_{1} + Z_{2}(p) \hat{v}_{2} \gamma^{2} \iu p_{2}  } (\rho^{0} \otimes \iu \tau^{1}),
    \end{equation}
    where $u \equiv \kappa^2 g^2 (N-1) /(16\pi^2)$.
    
    By collecting the terms of the same matrix structure, we find $\Sigma(q) = \gamma^{\mu} \Sigma_{\mu}(q)$ with
    \begin{subequations}\label{eq:Sigmas}
    \begin{align}
    		\Sigma_{0}(q) &=\int \frac{\D^3 p}{(2\pi)^{3}} \left[ +4\pi^2 u \frac{1}{\abs{p - q}} - \frac{e^2}{c^2} \mathcal{D}(p-q) \right] \frac{Z_{0} \iu p_{0} }{(Z_{0} p_{0})^2 + (Z_{1} \hat{v} p_{1})^2 + (Z_{2} \hat{v} p_{2})^2 } \\
    		\Sigma_{1}(q) &= \int \frac{\D^3 p}{(2\pi)^{3}} \left[ -4\pi^2 u \frac{1}{\abs{p - q}} + \frac{e^2}{c^2} \mathcal{D}(p-q) \right] \frac{Z_{1} \iu \hat{v}_1 p_{1} }{(Z_{0} p_{0})^2 + (Z_{1} \hat{v}p_{1})^2 + (Z_{2} \hat{v} p_{2})^2 } \\
    		\Sigma_{2}(q) &= \int \frac{\D^3 p}{(2\pi)^{3}} \left[ +4\pi^2 u \frac{1}{\abs{p - q}} + \frac{e^2}{c^2} \mathcal{D}(p-q) \right] \frac{Z_{2} \iu \hat{v}_2 p_{2} }{(Z_{0} p_{0})^2 + (Z_{1} \hat{v} p_{1})^2 + (Z_{2} \hat{v} p_{2})^2 }.
    \end{align}
    \end{subequations}
    Note that all spatial momenta in the above equations are rescaled by $1/c$.
    Moreover, note the different signs in front of the interaction kernels.
    They originate from $\gamma^{0} \gamma^{\mu} \gamma^{0} = \left( 2 \delta^{\mu 0} - 1 \right) \gamma^{\mu}$ and $ (\rho^{0} \otimes \iu \tau^{1}) \gamma^{\mu} (\rho^{0} \otimes \iu \tau^{1}) = \left( 1 - 2 \delta^{\mu 1} \right) \gamma^{\mu}$.
    Multiplying the DS equation by $\gamma^{\nu}$ and performing the matrix trace yields self-consistent equations for the renormalization factors $Z_{\mu}(q) = 1 + \delta Z_{\mu}(q)$.
    Working to leading order in $u$ and $1/N_s$, we can replace the dressed fermion propagators in the self-energies by the bare ones to find an approximate expression for $\delta Z_{\mu}(q)$.
    Realizing that the remaining integrals are linearly divergent, we expand the integrand in $q \ll p$ so that $\Sigma_{\mu}(q) = \iu \hat{v}_{\mu} q_{\mu} \delta Z_{\mu}(q)$, where $\delta Z_{\mu}(q)$ is the remaining log-divergent terms of Eqs.~\eqref{eq:Sigmas}.
    This yields the equations 
    \begin{subequations}
        \begin{align}
        	\delta Z_{0}(q) &= \int \frac{\D^3 p}{(2\pi)^{3}} \left[ +4\pi^2 u \frac{1}{\abs{p}} - \frac{e^2}{c^2}\mathcal{D}(p)  \right] \frac{\hat{v}^2 \v{p}^2 - (p_0)^2}{(p_0^2 + \hat{v}^2 \v{p}^2)^2} \\
        	\delta Z_{1}(q) &= \int \frac{\D^3 p}{(2\pi)^{3}} \left[ -4\pi^2 u \frac{1}{\abs{p}} + \frac{e^2}{c^2} \mathcal{D}(p)  \right] \frac{p_0^2 + \hat{v}^2 (p_2^2 - p_1^2)}{(p_0^2 + \hat{v}^2 \v{p}^2)^2} \\
        	\delta Z_{2}(q) &= \int \frac{\D^3 p}{(2\pi)^{3}} \left[ +4\pi^2 u \frac{1}{\abs{p}} + \frac{e^2}{c^2} \mathcal{D}(p)  \right] \frac{p_0^2 + \hat{v}^2 (p_1^2 - p_2^2)}{(p_0^2 + \hat{v}^2 \v{p}^2)^2}.
        \end{align} 
    \end{subequations}
    
    The integrals that multiply $u$ are of the form
    \begin{equation}
        J_{\mu}(\hat{v},q) = 	\int \frac{\D^3 p}{(2\pi)^3} \frac{p_{\mu}^2}{\sqrt{p^2} (p_{0}^2 + \hat{v}^2 \v{p}^2)^2}.
    \end{equation}
    Using a Feynman parameter $s$ and the change of variables $\ell_0 = p_0$, and $l_{i} = \sqrt{1 + s (\hat{v}^2 -1)} p_{i}$ 
    we get
    \begin{align}
        J_{\mu}(\hat{v},q) &=  \frac{\Gamma(5/2)}{\Gamma(1/2)} \int_{0}^{1} \D s \frac{s (1-s)^{-1/2}}{(1 + (\hat{v}^2 -1)s )} \int \frac{\D^3 \ell}{(2\pi)^3} \frac{p_{\mu}^2}{(\ell^2)^{5/2}} \notag \\
        &= \frac{3}{4} \int \frac{\D^3 \ell}{(2\pi)^3} \frac{\ell^2 /3}{(\ell^2)^{5/2}} \int_{0}^{1} \D s \frac{s (1-s)^{-1/2}}{(1 + (\hat{v}^2 - 1) s)^{\alpha}} = \frac{1}{8\pi^2} f_{\mu}(\hat{v}) \log \frac{\Lambda}{q},
    \end{align}
    where we have used spherical symmetry to replace $\ell_{\mu}^2 = \ell^2/3$ in the remaining momentum integral, which we ultimately perform as a momentum-shell integral.
    The power $\alpha = 1$ for $\mu = 0$ and $\alpha = 2$ for $\mu = i$.
    The functions $f_{\mu}(\hat{v})$ are found to be
    \begin{subequations}
    \begin{equation}\label{eq:f0}
    	f_{0}(\hat{v}) = \int_{0}^{1} \D s \frac{s (1-s)^{-1/2}}{1 + (\hat{v}^2 - 1) s} = 2 \left[\frac{1}{\hat{v}^2 - 1} + \frac{\arccos(\hat{v}) }{\hat{v} (1 - \hat{v}^2)^{3/2}} \right],
    \end{equation}
    and
    \begin{equation}\label{eq:f1}
    	f_{i}(\hat{v}) = \int_{0}^{1} \D s \frac{s (1-s)^{-1/2}}{(1 + (\hat{v}^2 - 1) s)^2}= \frac{1}{\hat{v}^2} \left[ \frac{1}{1-\hat{v}^2} - \frac{(2\hat{v}^2 - 1) \arccos(\hat{v})}{\hat{v} (1 - \hat{v}^2)^{3/2}} \right].
    \end{equation}
    \end{subequations}

    \begin{figure*}[t]
        \centering
        \includegraphics[width=\linewidth]{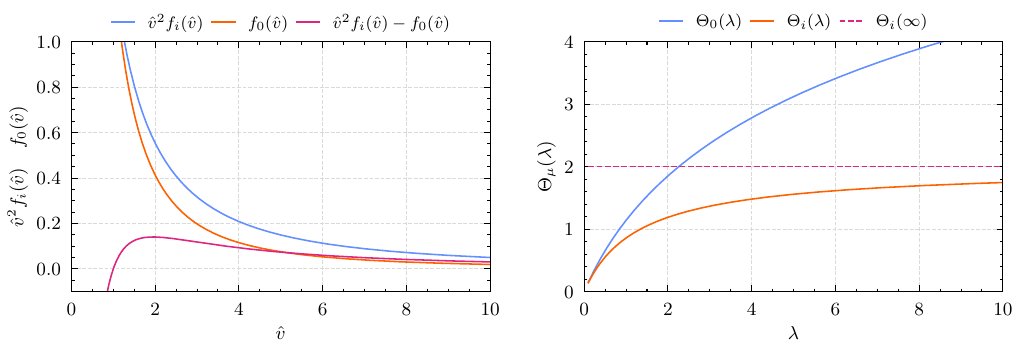}
        \caption{The functions $f_{0}(\hat{v})$, $\hat{v}^2 f_{i}(\hat{v})$ and their difference $\hat{v}^2 f_{i}(\hat{v}) - f_{0}(\hat{v})$ (left-hand side), as well as $\Theta_0(\lambda)$ and $\Theta_i(\lambda)$ together with the asymptotic value $\lim_{\lambda \to \infty} \Theta_i(\lambda) = 2$ (right-hand side).}
        \label{fig:fandTheta}
    \end{figure*}
    
    After a rescaling of the spatial momenta $\v{p} \mapsto \v{p}/\hat{v}$, the integrals that multiply $e^2$ can be written as 
    \begin{equation}
        I_{\mu}(\lambda,q) = \frac{1}{v^2} \int \frac{\D^3 p}{(2\pi)^3} \frac{p_{\mu}^2}{\abs{p}^4} \widetilde{e^2 \mathcal{D}(p)}.
    \end{equation}
    These are equivalent to the integrals appearing in the Appendix of Ref.~\cite{sonQuantumCriticalPoint2007}.
    By using spherical coordinates with $p_{0} = p \cos \vartheta$ and $\abs{\v{p}} = p \sin \vartheta$ and Eq.~\eqref{eq:Drescaled} one finds
    \begin{subequations}
        \begin{align}
    	I_0(\lambda,q) &= \frac{1}{8\pi^3} \frac{8 \lambda}{N_s}  2\pi \int_{0}^{\pi} \D \vartheta \sin\vartheta \frac{\cos^2\vartheta}{\sin\vartheta\left( 1 + \lambda \sin\vartheta\right)} \int_{q}^{\Lambda} \frac{\D p}{p} = \frac{2}{\pi^2 N_s} \Theta_{0}(\lambda) \log \frac{\Lambda}{q} ,
    \end{align}
    and
    \begin{align}
    	\sum_{i}I_i(\lambda,q) &= \frac{1}{8\pi^3} \frac{8 \lambda}{ N_s}  2\pi \int_{0}^{\pi} \D \vartheta \sin\vartheta \frac{\sin^2\vartheta}{\sin\vartheta\left( 1 + \lambda \sin\vartheta\right)} \int_{q}^{\Lambda} \frac{\D p}{p} = \frac{2}{\pi^2 N_s} \Theta_{i}(\lambda) \log \frac{\Lambda}{q} .
    \end{align}
    \end{subequations}
    The functions $\Theta_{0}(\lambda)$ and $\Theta_{i}(\lambda)$ are given by
    \begin{subequations}
        \begin{align}
        \Theta_{0}(\lambda) &=  -2 +\frac{\pi}{\lambda} + \frac{2\sqrt{\lambda^2 - 1}}{\lambda} \arcosh(\lambda),
        \intertext{and}
        \Theta_{i}(\lambda) &= 2 - \frac{\pi}{\lambda} + \frac{2}{\lambda \sqrt{\lambda^2 - 1}}\, \arcosh(\lambda). \label{eq:Theta} 
    \end{align}
    \end{subequations}
    Using $v_{i}(q) = v_i( 1 + \delta Z_{i}(q) - \delta Z_{0}(q) )$ one obtains Eq.~\eqref{eq:anomalous_dims}.
    The functions $f_{0}(\hat{v})$, $\hat{v}^2 f_{i}(\hat{v})$, $\hat{v}^2 f_{i}(\hat{v}) - f_{0}(\hat{v})$ and $\Theta_{\mu}(\lambda)$ are shown in Fig.~\ref{fig:fandTheta}.
    Since the scale that sets the fermion velocity is the hopping parameter, and the scale that sets the $\bm{\pi}$ boson velocity is the exchange coupling, we generically expect $\hat{v} \gtrsim 1$. 
    Taken together with the plot in Fig.~\ref{fig:fandTheta}, Eq.~\eqref{eq:anomalous_dims} indicates that the effect of magnetic fluctuations on $z$ is negligible compared to the Coulomb contribution.

    \section{Spontaneous mass generation}\label{app:mass_generation}

    The self-energy equation due to the magnetic fluctuations $\bm{\pi}$ in Eq.~\eqref{eq:Sigma_pi} can be modified to study spontaneous mass generation self-consistently (neglecting, for the moment, the long-range Coulomb interaction). 
    Indeed, replacing all wave-function renormalizations by $Z_{\mu} = 1$ and instead dressing the fermion propagator by a mass $\Sigma(p)$, we find a self-consistent equation for $\Sigma(p)$ which is simplified by taking the matrix trace of both sides
    \begin{align}
    	\Sigma(q) = \frac{\kappa^2 g^2 (N-1)}{4} \int \frac{\D^3 p}{(2\pi)^3} \frac{\Sigma(p)}{\Sigma^2(p) + p^2_{0} + \hat{v}^2 \v{p}^2} \frac{1}{\abs{p-q}}. 
    \end{align}
    Observe now that this is essentially the same equation that one gets from studying chiral symmetry breaking in QED$_3$ in the large-$N$ limit \cite{appelquistCriticalBehavior2+1Dimensional1988,appelquistSpontaneousChiralsymmetryBreaking1986}.
    The difference only lies in the coefficient.
    By setting $\hat{v}=1$ for simplicity, we can do the angular integrals and be left with
    \begin{equation}
        \Sigma(q) = \frac{u(N)}{q} \int_{0}^{\Lambda}\D p \frac{ p \Sigma(p)}{p^2 + \Sigma^2(p)} \left( p + q - \abs{p-q} \right),
    \end{equation}
    which can be converted into the differential equation
    \begin{equation}
    	\frac{\D }{\D q} \left( q^2 \frac{\D \Sigma(q)}{\D q} \right)  =  - 2u(N) \frac{q^2 \Sigma(q)}{q^2 + \Sigma^2(q) },
    \end{equation}
    with the boundary condition $0 \leq \Sigma(q) < \infty$ and 
    $
    	(q \D \Sigma(q) /\D q + \Sigma(q) ) \bigr\lvert_{q = \Lambda} = 0.
    $
    Following Ref.~\cite{appelquistCriticalBehavior2+1Dimensional1988} and linearizing the differential equation for $\Sigma(q)$, and using the \textit{ansatz} $\Sigma(q) \sim q^{\alpha}$ we get the condition
    \begin{equation}
    	\alpha(\alpha + 1) = - 2 u(N) \quad \Rightarrow \quad \alpha = - \frac{1}{2} \pm \sqrt{1 - 4 u(N)}.
    \end{equation}
    In this case, one can show that there is a critical $N$ \textit{beyond which} mass is generated spontaneously, which is given by $N_c = 1 + 4 \pi^2/(\kappa g)^2$.
    Being a perturbative result in $\kappa g$, the generically large $N_c$ leads us to conclude that the regime of spontaneous mass generation is unlikely to be relevant for this model. 
    Hence, the fluctuations of $\bm{\pi}$ do not destroy the nodal structure of the low-energy fermions. 
    However, due to the broken time-reversal symmetry, any other physical mechanism that can generate such a mass would be interesting since the resulting insulating state is possibly topologically nontrivial \cite{reDiracPointsTopological2024}.
    
    \twocolumngrid
    
	\bibliography{ref}

\begin{thebibliography}{65}%
\makeatletter
\providecommand \@ifxundefined [1]{%
 \@ifx{#1\undefined}
}%
\providecommand \@ifnum [1]{%
 \ifnum #1\expandafter \@firstoftwo
 \else \expandafter \@secondoftwo
 \fi
}%
\providecommand \@ifx [1]{%
 \ifx #1\expandafter \@firstoftwo
 \else \expandafter \@secondoftwo
 \fi
}%
\providecommand \natexlab [1]{#1}%
\providecommand \enquote  [1]{``#1''}%
\providecommand \bibnamefont  [1]{#1}%
\providecommand \bibfnamefont [1]{#1}%
\providecommand \citenamefont [1]{#1}%
\providecommand \href@noop [0]{\@secondoftwo}%
\providecommand \href [0]{\begingroup \@sanitize@url \@href}%
\providecommand \@href[1]{\@@startlink{#1}\@@href}%
\providecommand \@@href[1]{\endgroup#1\@@endlink}%
\providecommand \@sanitize@url [0]{\catcode `\\12\catcode `\$12\catcode `\&12\catcode `\#12\catcode `\^12\catcode `\_12\catcode `\%12\relax}%
\providecommand \@@startlink[1]{}%
\providecommand \@@endlink[0]{}%
\providecommand \url  [0]{\begingroup\@sanitize@url \@url }%
\providecommand \@url [1]{\endgroup\@href {#1}{\urlprefix }}%
\providecommand \urlprefix  [0]{URL }%
\providecommand \Eprint [0]{\href }%
\providecommand \doibase [0]{https://doi.org/}%
\providecommand \selectlanguage [0]{\@gobble}%
\providecommand \bibinfo  [0]{\@secondoftwo}%
\providecommand \bibfield  [0]{\@secondoftwo}%
\providecommand \translation [1]{[#1]}%
\providecommand \BibitemOpen [0]{}%
\providecommand \bibitemStop [0]{}%
\providecommand \bibitemNoStop [0]{.\EOS\space}%
\providecommand \EOS [0]{\spacefactor3000\relax}%
\providecommand \BibitemShut  [1]{\csname bibitem#1\endcsname}%
\let\auto@bib@innerbib\@empty
\bibitem [{\citenamefont {Chakravarty}\ \emph {et~al.}(1988)\citenamefont {Chakravarty}, \citenamefont {Halperin},\ and\ \citenamefont {Nelson}}]{chakravartyLowtemperatureBehaviorTwodimensional1988}%
  \BibitemOpen
  \bibfield  {author} {\bibinfo {author} {\bibfnamefont {S.}~\bibnamefont {Chakravarty}}, \bibinfo {author} {\bibfnamefont {B.~I.}\ \bibnamefont {Halperin}},\ and\ \bibinfo {author} {\bibfnamefont {D.~R.}\ \bibnamefont {Nelson}},\ }\bibfield  {title} {\bibinfo {title} {Low-temperature behavior of two-dimensional quantum antiferromagnets},\ }\href {https://doi.org/10.1103/PhysRevLett.60.1057} {\bibfield  {journal} {\bibinfo  {journal} {Phys. Rev. Lett.}\ }\textbf {\bibinfo {volume} {60}},\ \bibinfo {pages} {1057} (\bibinfo {year} {1988})}\BibitemShut {NoStop}%
\bibitem [{\citenamefont {Chakravarty}\ \emph {et~al.}(1989)\citenamefont {Chakravarty}, \citenamefont {Halperin},\ and\ \citenamefont {Nelson}}]{chakravartyTwodimensionalQuantumHeisenberg1989}%
  \BibitemOpen
  \bibfield  {author} {\bibinfo {author} {\bibfnamefont {S.}~\bibnamefont {Chakravarty}}, \bibinfo {author} {\bibfnamefont {B.~I.}\ \bibnamefont {Halperin}},\ and\ \bibinfo {author} {\bibfnamefont {D.~R.}\ \bibnamefont {Nelson}},\ }\bibfield  {title} {\bibinfo {title} {Two-dimensional quantum {{Heisenberg}} antiferromagnet at low temperatures},\ }\href {https://doi.org/10.1103/PhysRevB.39.2344} {\bibfield  {journal} {\bibinfo  {journal} {Phys. Rev. B}\ }\textbf {\bibinfo {volume} {39}},\ \bibinfo {pages} {2344} (\bibinfo {year} {1989})}\BibitemShut {NoStop}%
\bibitem [{\citenamefont {Auerbach}(1994)}]{auerbachInteractingElectronsQuantum1994}%
  \BibitemOpen
  \bibfield  {author} {\bibinfo {author} {\bibfnamefont {A.}~\bibnamefont {Auerbach}},\ }\href {https://doi.org/10.1007/978-1-4612-0869-3} {\emph {\bibinfo {title} {Interacting {{Electrons}} and {{Quantum Magnetism}}}}},\ Graduate {{Texts}} in {{Contemporary Physics}}\ (\bibinfo  {publisher} {Springer New York},\ \bibinfo {address} {New York, NY},\ \bibinfo {year} {1994})\BibitemShut {NoStop}%
\bibitem [{\citenamefont {Anderson}(1987)}]{andersonResonatingValenceBond1987}%
  \BibitemOpen
  \bibfield  {author} {\bibinfo {author} {\bibfnamefont {P.~W.}\ \bibnamefont {Anderson}},\ }\bibfield  {title} {\bibinfo {title} {The {{Resonating Valence Bond State}} in {{La2CuO4}} and {{Superconductivity}}},\ }\href {https://doi.org/10.1126/science.235.4793.1196} {\bibfield  {journal} {\bibinfo  {journal} {Science}\ }\textbf {\bibinfo {volume} {235}},\ \bibinfo {pages} {1196} (\bibinfo {year} {1987})}\BibitemShut {NoStop}%
\bibitem [{\citenamefont {Affleck}\ \emph {et~al.}(1988)\citenamefont {Affleck}, \citenamefont {Zou}, \citenamefont {Hsu},\ and\ \citenamefont {Anderson}}]{affleckSU2GaugeSymmetry1988}%
  \BibitemOpen
  \bibfield  {author} {\bibinfo {author} {\bibfnamefont {I.}~\bibnamefont {Affleck}}, \bibinfo {author} {\bibfnamefont {Z.}~\bibnamefont {Zou}}, \bibinfo {author} {\bibfnamefont {T.}~\bibnamefont {Hsu}},\ and\ \bibinfo {author} {\bibfnamefont {P.~W.}\ \bibnamefont {Anderson}},\ }\bibfield  {title} {\bibinfo {title} {{SU(2) gauge symmetry of the large-$U$ limit of the Hubbard model}},\ }\href {https://doi.org/10.1103/PhysRevB.38.745} {\bibfield  {journal} {\bibinfo  {journal} {Phys. Rev. B}\ }\textbf {\bibinfo {volume} {38}},\ \bibinfo {pages} {745} (\bibinfo {year} {1988})}\BibitemShut {NoStop}%
\bibitem [{\citenamefont {Baltz}\ \emph {et~al.}(2018)\citenamefont {Baltz}, \citenamefont {Manchon}, \citenamefont {Tsoi}, \citenamefont {Moriyama}, \citenamefont {Ono},\ and\ \citenamefont {Tserkovnyak}}]{baltzAntiferromagneticSpintronics2018}%
  \BibitemOpen
  \bibfield  {author} {\bibinfo {author} {\bibfnamefont {V.}~\bibnamefont {Baltz}}, \bibinfo {author} {\bibfnamefont {A.}~\bibnamefont {Manchon}}, \bibinfo {author} {\bibfnamefont {M.}~\bibnamefont {Tsoi}}, \bibinfo {author} {\bibfnamefont {T.}~\bibnamefont {Moriyama}}, \bibinfo {author} {\bibfnamefont {T.}~\bibnamefont {Ono}},\ and\ \bibinfo {author} {\bibfnamefont {Y.}~\bibnamefont {Tserkovnyak}},\ }\bibfield  {title} {\bibinfo {title} {Antiferromagnetic spintronics},\ }\href {https://doi.org/10.1103/RevModPhys.90.015005} {\bibfield  {journal} {\bibinfo  {journal} {Rev. Mod. Phys.}\ }\textbf {\bibinfo {volume} {90}},\ \bibinfo {pages} {015005} (\bibinfo {year} {2018})}\BibitemShut {NoStop}%
\bibitem [{\citenamefont {Sachdev}\ and\ \citenamefont {Jalabert}(1990)}]{sachdevEffectiveLatticeModels1990}%
  \BibitemOpen
  \bibfield  {author} {\bibinfo {author} {\bibfnamefont {S.}~\bibnamefont {Sachdev}}\ and\ \bibinfo {author} {\bibfnamefont {R.}~\bibnamefont {Jalabert}},\ }\bibfield  {title} {\bibinfo {title} {Effective lattice models for two-dimensional quantum antiferromagnets},\ }\href {https://doi.org/10.1142/S0217984990001318} {\bibfield  {journal} {\bibinfo  {journal} {Mod. Phys. Lett. B}\ }\textbf {\bibinfo {volume} {04}},\ \bibinfo {pages} {1043} (\bibinfo {year} {1990})}\BibitemShut {NoStop}%
\bibitem [{\citenamefont {Mudry}\ and\ \citenamefont {Fradkin}(1994)}]{mudrySeparationSpinCharge1994}%
  \BibitemOpen
  \bibfield  {author} {\bibinfo {author} {\bibfnamefont {C.}~\bibnamefont {Mudry}}\ and\ \bibinfo {author} {\bibfnamefont {E.}~\bibnamefont {Fradkin}},\ }\bibfield  {title} {\bibinfo {title} {Separation of spin and charge quantum numbers in strongly correlated systems},\ }\href {https://doi.org/10.1103/PhysRevB.49.5200} {\bibfield  {journal} {\bibinfo  {journal} {Phys. Rev. B}\ }\textbf {\bibinfo {volume} {49}},\ \bibinfo {pages} {5200} (\bibinfo {year} {1994})}\BibitemShut {NoStop}%
\bibitem [{\citenamefont {Wen}(2002)}]{wenQuantumOrdersSymmetric2002}%
  \BibitemOpen
  \bibfield  {author} {\bibinfo {author} {\bibfnamefont {X.-G.}\ \bibnamefont {Wen}},\ }\bibfield  {title} {\bibinfo {title} {Quantum orders and symmetric spin liquids},\ }\href {https://doi.org/10.1103/PhysRevB.65.165113} {\bibfield  {journal} {\bibinfo  {journal} {Phys. Rev. B}\ }\textbf {\bibinfo {volume} {65}},\ \bibinfo {pages} {165113} (\bibinfo {year} {2002})}\BibitemShut {NoStop}%
\bibitem [{\citenamefont {Kosterlitz}\ and\ \citenamefont {Thouless}(1973)}]{kosterlitzOrderingMetastabilityPhase1973}%
  \BibitemOpen
  \bibfield  {author} {\bibinfo {author} {\bibfnamefont {J.~M.}\ \bibnamefont {Kosterlitz}}\ and\ \bibinfo {author} {\bibfnamefont {D.~J.}\ \bibnamefont {Thouless}},\ }\bibfield  {title} {\bibinfo {title} {Ordering, metastability and phase transitions in two-dimensional systems},\ }\href {https://doi.org/10.1088/0022-3719/6/7/010} {\bibfield  {journal} {\bibinfo  {journal} {J. Phys. C: Solid State Phys.}\ }\textbf {\bibinfo {volume} {6}},\ \bibinfo {pages} {1181} (\bibinfo {year} {1973})}\BibitemShut {NoStop}%
\bibitem [{\citenamefont {Nelson}\ and\ \citenamefont {Kosterlitz}(1977)}]{nelsonUniversalJumpSuperfluid1977}%
  \BibitemOpen
  \bibfield  {author} {\bibinfo {author} {\bibfnamefont {D.~R.}\ \bibnamefont {Nelson}}\ and\ \bibinfo {author} {\bibfnamefont {J.~M.}\ \bibnamefont {Kosterlitz}},\ }\bibfield  {title} {\bibinfo {title} {Universal {{Jump}} in the {{Superfluid Density}} of {{Two-Dimensional Superfluids}}},\ }\href {https://doi.org/10.1103/PhysRevLett.39.1201} {\bibfield  {journal} {\bibinfo  {journal} {Phys. Rev. Lett.}\ }\textbf {\bibinfo {volume} {39}},\ \bibinfo {pages} {1201} (\bibinfo {year} {1977})}\BibitemShut {NoStop}%
\bibitem [{\citenamefont {Haldane}(1988)}]{haldaneO3Nonlinear$ensuremathsigma$1988}%
  \BibitemOpen
  \bibfield  {author} {\bibinfo {author} {\bibfnamefont {F.~D.~M.}\ \bibnamefont {Haldane}},\ }\bibfield  {title} {\bibinfo {title} {O(3) nonlinear $\ensuremath{\sigma}$ model and the topological distinction between integer- and half-integer-spin antiferromagnets in two dimensions},\ }\href {https://doi.org/10.1103/PhysRevLett.61.1029} {\bibfield  {journal} {\bibinfo  {journal} {Phys. Rev. Lett.}\ }\textbf {\bibinfo {volume} {61}},\ \bibinfo {pages} {1029} (\bibinfo {year} {1988})}\BibitemShut {NoStop}%
\bibitem [{\citenamefont {Read}\ and\ \citenamefont {Sachdev}(1989)}]{readValencebondSpinPeierlsGround1989}%
  \BibitemOpen
  \bibfield  {author} {\bibinfo {author} {\bibfnamefont {N.}~\bibnamefont {Read}}\ and\ \bibinfo {author} {\bibfnamefont {S.}~\bibnamefont {Sachdev}},\ }\bibfield  {title} {\bibinfo {title} {Valence-bond and spin-{{Peierls}} ground states of low-dimensional quantum antiferromagnets},\ }\href {https://doi.org/10.1103/PhysRevLett.62.1694} {\bibfield  {journal} {\bibinfo  {journal} {Phys. Rev. Lett.}\ }\textbf {\bibinfo {volume} {62}},\ \bibinfo {pages} {1694} (\bibinfo {year} {1989})}\BibitemShut {NoStop}%
\bibitem [{\citenamefont {Senthil}\ \emph {et~al.}(2004{\natexlab{a}})\citenamefont {Senthil}, \citenamefont {Vishwanath}, \citenamefont {Balents}, \citenamefont {Sachdev},\ and\ \citenamefont {Fisher}}]{senthilDeconfinedQuantumCritical2004}%
  \BibitemOpen
  \bibfield  {author} {\bibinfo {author} {\bibfnamefont {T.}~\bibnamefont {Senthil}}, \bibinfo {author} {\bibfnamefont {A.}~\bibnamefont {Vishwanath}}, \bibinfo {author} {\bibfnamefont {L.}~\bibnamefont {Balents}}, \bibinfo {author} {\bibfnamefont {S.}~\bibnamefont {Sachdev}},\ and\ \bibinfo {author} {\bibfnamefont {M.~P.~A.}\ \bibnamefont {Fisher}},\ }\bibfield  {title} {\bibinfo {title} {Deconfined {{Quantum Critical Points}}},\ }\href {https://doi.org/10.1126/science.1091806} {\bibfield  {journal} {\bibinfo  {journal} {Science}\ }\textbf {\bibinfo {volume} {303}},\ \bibinfo {pages} {1490} (\bibinfo {year} {2004}{\natexlab{a}})}\BibitemShut {NoStop}%
\bibitem [{\citenamefont {Senthil}\ \emph {et~al.}(2004{\natexlab{b}})\citenamefont {Senthil}, \citenamefont {Balents}, \citenamefont {Sachdev}, \citenamefont {Vishwanath},\ and\ \citenamefont {Fisher}}]{senthilQuantumCriticalityLandauGinzburgWilson2004}%
  \BibitemOpen
  \bibfield  {author} {\bibinfo {author} {\bibfnamefont {T.}~\bibnamefont {Senthil}}, \bibinfo {author} {\bibfnamefont {L.}~\bibnamefont {Balents}}, \bibinfo {author} {\bibfnamefont {S.}~\bibnamefont {Sachdev}}, \bibinfo {author} {\bibfnamefont {A.}~\bibnamefont {Vishwanath}},\ and\ \bibinfo {author} {\bibfnamefont {M.~P.~A.}\ \bibnamefont {Fisher}},\ }\bibfield  {title} {\bibinfo {title} {Quantum criticality beyond the {{Landau-Ginzburg-Wilson}} paradigm},\ }\href {https://doi.org/10.1103/PhysRevB.70.144407} {\bibfield  {journal} {\bibinfo  {journal} {Phys. Rev. B}\ }\textbf {\bibinfo {volume} {70}},\ \bibinfo {pages} {144407} (\bibinfo {year} {2004}{\natexlab{b}})}\BibitemShut {NoStop}%
\bibitem [{\citenamefont {Senthil}\ \emph {et~al.}(2005)\citenamefont {Senthil}, \citenamefont {Balents}, \citenamefont {Sachdev}, \citenamefont {Vishwanath},\ and\ \citenamefont {P.~A.~Fisher}}]{senthilDeconfinedCriticalityCritically2005}%
  \BibitemOpen
  \bibfield  {author} {\bibinfo {author} {\bibfnamefont {T.}~\bibnamefont {Senthil}}, \bibinfo {author} {\bibfnamefont {L.}~\bibnamefont {Balents}}, \bibinfo {author} {\bibfnamefont {S.}~\bibnamefont {Sachdev}}, \bibinfo {author} {\bibfnamefont {A.}~\bibnamefont {Vishwanath}},\ and\ \bibinfo {author} {\bibfnamefont {M.}~\bibnamefont {P.~A.~Fisher}},\ }\bibfield  {title} {\bibinfo {title} {Deconfined {{Criticality Critically Defined}}},\ }\href {https://doi.org/10.1143/JPSJS.74S.1} {\bibfield  {journal} {\bibinfo  {journal} {J. Phys. Soc. Jpn.}\ }\textbf {\bibinfo {volume} {74}},\ \bibinfo {pages} {1} (\bibinfo {year} {2005})}\BibitemShut {NoStop}%
\bibitem [{\citenamefont {Kragset}\ \emph {et~al.}(2006)\citenamefont {Kragset}, \citenamefont {Sm{\o}rgrav}, \citenamefont {Hove}, \citenamefont {Nogueira},\ and\ \citenamefont {Sudb{\o}}}]{kragsetFirstOrderPhaseTransition2006}%
  \BibitemOpen
  \bibfield  {author} {\bibinfo {author} {\bibfnamefont {S.}~\bibnamefont {Kragset}}, \bibinfo {author} {\bibfnamefont {E.}~\bibnamefont {Sm{\o}rgrav}}, \bibinfo {author} {\bibfnamefont {J.}~\bibnamefont {Hove}}, \bibinfo {author} {\bibfnamefont {F.~S.}\ \bibnamefont {Nogueira}},\ and\ \bibinfo {author} {\bibfnamefont {A.}~\bibnamefont {Sudb{\o}}},\ }\bibfield  {title} {\bibinfo {title} {First-{{Order Phase Transition}} in {{Easy-Plane Quantum Antiferromagnets}}},\ }\href {https://doi.org/10.1103/PhysRevLett.97.247201} {\bibfield  {journal} {\bibinfo  {journal} {Phys. Rev. Lett.}\ }\textbf {\bibinfo {volume} {97}},\ \bibinfo {pages} {247201} (\bibinfo {year} {2006})}\BibitemShut {NoStop}%
\bibitem [{\citenamefont {Kuklov}\ \emph {et~al.}(2006)\citenamefont {Kuklov}, \citenamefont {Prokof'ev}, \citenamefont {Svistunov},\ and\ \citenamefont {Troyer}}]{kuklovDeconfinedCriticalityRunaway2006}%
  \BibitemOpen
  \bibfield  {author} {\bibinfo {author} {\bibfnamefont {A.~B.}\ \bibnamefont {Kuklov}}, \bibinfo {author} {\bibfnamefont {N.~V.}\ \bibnamefont {Prokof'ev}}, \bibinfo {author} {\bibfnamefont {B.~V.}\ \bibnamefont {Svistunov}},\ and\ \bibinfo {author} {\bibfnamefont {M.}~\bibnamefont {Troyer}},\ }\bibfield  {title} {\bibinfo {title} {Deconfined criticality, runaway flow in the two-component scalar electrodynamics and weak first-order superfluid-solid transitions},\ }\href {https://doi.org/10.1016/j.aop.2006.04.007} {\bibfield  {journal} {\bibinfo  {journal} {Annals of Physics}\ }\bibinfo {series} {July 2006 {{Special Issue}}},\ \textbf {\bibinfo {volume} {321}},\ \bibinfo {pages} {1602} (\bibinfo {year} {2006})}\BibitemShut {NoStop}%
\bibitem [{\citenamefont {Kuklov}\ \emph {et~al.}(2008)\citenamefont {Kuklov}, \citenamefont {Matsumoto}, \citenamefont {Prokof'ev}, \citenamefont {Svistunov},\ and\ \citenamefont {Troyer}}]{kuklovDeconfinedCriticalityGeneric2008}%
  \BibitemOpen
  \bibfield  {author} {\bibinfo {author} {\bibfnamefont {A.~B.}\ \bibnamefont {Kuklov}}, \bibinfo {author} {\bibfnamefont {M.}~\bibnamefont {Matsumoto}}, \bibinfo {author} {\bibfnamefont {N.~V.}\ \bibnamefont {Prokof'ev}}, \bibinfo {author} {\bibfnamefont {B.~V.}\ \bibnamefont {Svistunov}},\ and\ \bibinfo {author} {\bibfnamefont {M.}~\bibnamefont {Troyer}},\ }\bibfield  {title} {\bibinfo {title} {Deconfined {{Criticality}}: {{Generic First-Order Transition}} in the {{SU}}(2) {{Symmetry Case}}},\ }\href {https://doi.org/10.1103/PhysRevLett.101.050405} {\bibfield  {journal} {\bibinfo  {journal} {Phys. Rev. Lett.}\ }\textbf {\bibinfo {volume} {101}},\ \bibinfo {pages} {050405} (\bibinfo {year} {2008})}\BibitemShut {NoStop}%
\bibitem [{\citenamefont {Herland}\ \emph {et~al.}(2013)\citenamefont {Herland}, \citenamefont {Bojesen}, \citenamefont {Babaev},\ and\ \citenamefont {Sudbø}}]{herlandPhaseStructurePhase2013}%
  \BibitemOpen
  \bibfield  {author} {\bibinfo {author} {\bibfnamefont {E.~V.}\ \bibnamefont {Herland}}, \bibinfo {author} {\bibfnamefont {T.~A.}\ \bibnamefont {Bojesen}}, \bibinfo {author} {\bibfnamefont {E.}~\bibnamefont {Babaev}},\ and\ \bibinfo {author} {\bibfnamefont {A.}~\bibnamefont {Sudbø}},\ }\bibfield  {title} {\bibinfo {title} {Phase structure and phase transitions in a three-dimensional {$SU(2)$} superconductor},\ }\href {https://doi.org/10.1103/PhysRevB.87.134503} {\bibfield  {journal} {\bibinfo  {journal} {Phys. Rev. B}\ }\textbf {\bibinfo {volume} {87}},\ \bibinfo {pages} {134503} (\bibinfo {year} {2013})}\BibitemShut {NoStop}%
\bibitem [{\citenamefont {{$\check{\text{S}}$}mejkal}\ \emph {et~al.}(2022{\natexlab{a}})\citenamefont {{$\check{\text{S}}$}mejkal}, \citenamefont {Sinova},\ and\ \citenamefont {Jungwirth}}]{smejkalEmergingResearchLandscape2022}%
  \BibitemOpen
  \bibfield  {author} {\bibinfo {author} {\bibfnamefont {L.}~\bibnamefont {{$\check{\text{S}}$}mejkal}}, \bibinfo {author} {\bibfnamefont {J.}~\bibnamefont {Sinova}},\ and\ \bibinfo {author} {\bibfnamefont {T.}~\bibnamefont {Jungwirth}},\ }\bibfield  {title} {\bibinfo {title} {Emerging {{Research Landscape}} of {{Altermagnetism}}},\ }\href {https://doi.org/10.1103/PhysRevX.12.040501} {\bibfield  {journal} {\bibinfo  {journal} {Phys. Rev. X}\ }\textbf {\bibinfo {volume} {12}},\ \bibinfo {pages} {040501} (\bibinfo {year} {2022}{\natexlab{a}})}\BibitemShut {NoStop}%
\bibitem [{\citenamefont {{$\check{\text{S}}$}mejkal}\ \emph {et~al.}(2022{\natexlab{b}})\citenamefont {{$\check{\text{S}}$}mejkal}, \citenamefont {Sinova},\ and\ \citenamefont {Jungwirth}}]{smejkalConventionalFerromagnetismAntiferromagnetism2022}%
  \BibitemOpen
  \bibfield  {author} {\bibinfo {author} {\bibfnamefont {L.}~\bibnamefont {{$\check{\text{S}}$}mejkal}}, \bibinfo {author} {\bibfnamefont {J.}~\bibnamefont {Sinova}},\ and\ \bibinfo {author} {\bibfnamefont {T.}~\bibnamefont {Jungwirth}},\ }\bibfield  {title} {\bibinfo {title} {Beyond {{Conventional Ferromagnetism}} and {{Antiferromagnetism}}: {{A Phase}} with {{Nonrelativistic Spin}} and {{Crystal Rotation Symmetry}}},\ }\href {https://doi.org/10.1103/PhysRevX.12.031042} {\bibfield  {journal} {\bibinfo  {journal} {Phys. Rev. X}\ }\textbf {\bibinfo {volume} {12}},\ \bibinfo {pages} {031042} (\bibinfo {year} {2022}{\natexlab{b}})}\BibitemShut {NoStop}%
\bibitem [{\citenamefont {Brekke}\ \emph {et~al.}(2023)\citenamefont {Brekke}, \citenamefont {Brataas},\ and\ \citenamefont {Sudb{\o}}}]{brekkeTwodimensionalAltermagnetsSuperconductivity2023}%
  \BibitemOpen
  \bibfield  {author} {\bibinfo {author} {\bibfnamefont {B.}~\bibnamefont {Brekke}}, \bibinfo {author} {\bibfnamefont {A.}~\bibnamefont {Brataas}},\ and\ \bibinfo {author} {\bibfnamefont {A.}~\bibnamefont {Sudb{\o}}},\ }\bibfield  {title} {\bibinfo {title} {Two-dimensional altermagnets: {{Superconductivity}} in a minimal microscopic model},\ }\href {https://doi.org/10.1103/PhysRevB.108.224421} {\bibfield  {journal} {\bibinfo  {journal} {Phys. Rev. B}\ }\textbf {\bibinfo {volume} {108}},\ \bibinfo {pages} {224421} (\bibinfo {year} {2023})}\BibitemShut {NoStop}%
\bibitem [{\citenamefont {M\ae{}land}\ \emph {et~al.}(2024)\citenamefont {M\ae{}land}, \citenamefont {Brekke},\ and\ \citenamefont {Sudb\o{}}}]{maeland_2024}%
  \BibitemOpen
  \bibfield  {author} {\bibinfo {author} {\bibfnamefont {K.}~\bibnamefont {M\ae{}land}}, \bibinfo {author} {\bibfnamefont {B.}~\bibnamefont {Brekke}},\ and\ \bibinfo {author} {\bibfnamefont {A.}~\bibnamefont {Sudb\o{}}},\ }\bibfield  {title} {\bibinfo {title} {Many-body effects on superconductivity mediated by double-magnon processes in altermagnets},\ }\href {https://doi.org/10.1103/PhysRevB.109.134515} {\bibfield  {journal} {\bibinfo  {journal} {Phys. Rev. B}\ }\textbf {\bibinfo {volume} {109}},\ \bibinfo {pages} {134515} (\bibinfo {year} {2024})}\BibitemShut {NoStop}%
\bibitem [{\citenamefont {Roig}\ \emph {et~al.}(2024)\citenamefont {Roig}, \citenamefont {Kreisel}, \citenamefont {Yu}, \citenamefont {Andersen},\ and\ \citenamefont {Agterberg}}]{roigMinimalModelsAltermagnetism2024}%
  \BibitemOpen
  \bibfield  {author} {\bibinfo {author} {\bibfnamefont {M.}~\bibnamefont {Roig}}, \bibinfo {author} {\bibfnamefont {A.}~\bibnamefont {Kreisel}}, \bibinfo {author} {\bibfnamefont {Y.}~\bibnamefont {Yu}}, \bibinfo {author} {\bibfnamefont {B.~M.}\ \bibnamefont {Andersen}},\ and\ \bibinfo {author} {\bibfnamefont {D.~F.}\ \bibnamefont {Agterberg}},\ }\bibfield  {title} {\bibinfo {title} {Minimal models for altermagnetism},\ }\href {https://doi.org/10.1103/PhysRevB.110.144412} {\bibfield  {journal} {\bibinfo  {journal} {Phys. Rev. B}\ }\textbf {\bibinfo {volume} {110}},\ \bibinfo {pages} {144412} (\bibinfo {year} {2024})}\BibitemShut {NoStop}%
\bibitem [{\citenamefont {Lee}\ \emph {et~al.}(2024)\citenamefont {Lee}, \citenamefont {Lee}, \citenamefont {Jung}, \citenamefont {Jung}, \citenamefont {Kim}, \citenamefont {Lee}, \citenamefont {Seok}, \citenamefont {Kim}, \citenamefont {Park}, \citenamefont {{$\check{\text{S}}$}mejkal}, \citenamefont {Kang},\ and\ \citenamefont {Kim}}]{leeBrokenKramersDegeneracy2024}%
  \BibitemOpen
  \bibfield  {author} {\bibinfo {author} {\bibfnamefont {S.}~\bibnamefont {Lee}}, \bibinfo {author} {\bibfnamefont {S.}~\bibnamefont {Lee}}, \bibinfo {author} {\bibfnamefont {S.}~\bibnamefont {Jung}}, \bibinfo {author} {\bibfnamefont {J.}~\bibnamefont {Jung}}, \bibinfo {author} {\bibfnamefont {D.}~\bibnamefont {Kim}}, \bibinfo {author} {\bibfnamefont {Y.}~\bibnamefont {Lee}}, \bibinfo {author} {\bibfnamefont {B.}~\bibnamefont {Seok}}, \bibinfo {author} {\bibfnamefont {J.}~\bibnamefont {Kim}}, \bibinfo {author} {\bibfnamefont {B.~G.}\ \bibnamefont {Park}}, \bibinfo {author} {\bibfnamefont {L.}~\bibnamefont {{$\check{\text{S}}$}mejkal}}, \bibinfo {author} {\bibfnamefont {C.-J.}\ \bibnamefont {Kang}},\ and\ \bibinfo {author} {\bibfnamefont {C.}~\bibnamefont {Kim}},\ }\bibfield  {title} {\bibinfo {title} {Broken {{Kramers Degeneracy}} in {{Altermagnetic MnTe}}},\ }\href {https://doi.org/10.1103/PhysRevLett.132.036702} {\bibfield  {journal} {\bibinfo  {journal} {Phys. Rev. Lett.}\ }\textbf {\bibinfo {volume}
  {132}},\ \bibinfo {pages} {036702} (\bibinfo {year} {2024})}\BibitemShut {NoStop}%
\bibitem [{\citenamefont {Krempask{\'y}}\ \emph {et~al.}(2024)\citenamefont {Krempask{\'y}}, \citenamefont {{$\check{\text{S}}$}mejkal}, \citenamefont {D'Souza}, \citenamefont {Hajlaoui}, \citenamefont {Springholz}, \citenamefont {Uhl{\'i}{$\check{\text{r}}$}ov{\'a}}, \citenamefont {Alarab}, \citenamefont {Constantinou}, \citenamefont {Strocov}, \citenamefont {Usanov}, \citenamefont {Pudelko}, \citenamefont {{Gonz{\'a}lez-Hern{\'a}ndez}}, \citenamefont {Birk~Hellenes}, \citenamefont {Jansa}, \citenamefont {Reichlov{\'a}}, \citenamefont {{$\check{\text{S}}$}ob{\'a}{$\check{\text{n}}$}}, \citenamefont {Gonzalez~Betancourt}, \citenamefont {Wadley}, \citenamefont {Sinova}, \citenamefont {Kriegner}, \citenamefont {Min{\'a}r}, \citenamefont {Dil},\ and\ \citenamefont {Jungwirth}}]{krempaskyAltermagneticLiftingKramers2024}%
  \BibitemOpen
  \bibfield  {author} {\bibinfo {author} {\bibfnamefont {J.}~\bibnamefont {Krempask{\'y}}}, \bibinfo {author} {\bibfnamefont {L.}~\bibnamefont {{$\check{\text{S}}$}mejkal}}, \bibinfo {author} {\bibfnamefont {S.~W.}\ \bibnamefont {D'Souza}}, \bibinfo {author} {\bibfnamefont {M.}~\bibnamefont {Hajlaoui}}, \bibinfo {author} {\bibfnamefont {G.}~\bibnamefont {Springholz}}, \bibinfo {author} {\bibfnamefont {K.}~\bibnamefont {Uhl{\'i}{$\check{\text{r}}$}ov{\'a}}}, \bibinfo {author} {\bibfnamefont {F.}~\bibnamefont {Alarab}}, \bibinfo {author} {\bibfnamefont {P.~C.}\ \bibnamefont {Constantinou}}, \bibinfo {author} {\bibfnamefont {V.}~\bibnamefont {Strocov}}, \bibinfo {author} {\bibfnamefont {D.}~\bibnamefont {Usanov}}, \bibinfo {author} {\bibfnamefont {W.~R.}\ \bibnamefont {Pudelko}}, \bibinfo {author} {\bibfnamefont {R.}~\bibnamefont {{Gonz{\'a}lez-Hern{\'a}ndez}}}, \bibinfo {author} {\bibfnamefont {A.}~\bibnamefont {Birk~Hellenes}}, \bibinfo {author} {\bibfnamefont {Z.}~\bibnamefont {Jansa}}, \bibinfo {author}
  {\bibfnamefont {H.}~\bibnamefont {Reichlov{\'a}}}, \bibinfo {author} {\bibfnamefont {Z.}~\bibnamefont {{$\check{\text{S}}$}ob{\'a}{$\check{\text{n}}$}}}, \bibinfo {author} {\bibfnamefont {R.~D.}\ \bibnamefont {Gonzalez~Betancourt}}, \bibinfo {author} {\bibfnamefont {P.}~\bibnamefont {Wadley}}, \bibinfo {author} {\bibfnamefont {J.}~\bibnamefont {Sinova}}, \bibinfo {author} {\bibfnamefont {D.}~\bibnamefont {Kriegner}}, \bibinfo {author} {\bibfnamefont {J.}~\bibnamefont {Min{\'a}r}}, \bibinfo {author} {\bibfnamefont {J.~H.}\ \bibnamefont {Dil}},\ and\ \bibinfo {author} {\bibfnamefont {T.}~\bibnamefont {Jungwirth}},\ }\bibfield  {title} {\bibinfo {title} {Altermagnetic lifting of {{Kramers}} spin degeneracy},\ }\href {https://doi.org/10.1038/s41586-023-06907-7} {\bibfield  {journal} {\bibinfo  {journal} {Nature}\ }\textbf {\bibinfo {volume} {626}},\ \bibinfo {pages} {517} (\bibinfo {year} {2024})}\BibitemShut {NoStop}%
\bibitem [{\citenamefont {Reimers}\ \emph {et~al.}(2024)\citenamefont {Reimers}, \citenamefont {Odenbreit}, \citenamefont {{$\check{\text{S}}$}mejkal}, \citenamefont {Strocov}, \citenamefont {Constantinou}, \citenamefont {Hellenes}, \citenamefont {Jaeschke~Ubiergo}, \citenamefont {Campos}, \citenamefont {Bharadwaj}, \citenamefont {Chakraborty}, \citenamefont {Denneulin}, \citenamefont {Shi}, \citenamefont {{Dunin-Borkowski}}, \citenamefont {Das}, \citenamefont {Kl{\"a}ui}, \citenamefont {Sinova},\ and\ \citenamefont {Jourdan}}]{reimersDirectObservationAltermagnetic2024}%
  \BibitemOpen
  \bibfield  {author} {\bibinfo {author} {\bibfnamefont {S.}~\bibnamefont {Reimers}}, \bibinfo {author} {\bibfnamefont {L.}~\bibnamefont {Odenbreit}}, \bibinfo {author} {\bibfnamefont {L.}~\bibnamefont {{$\check{\text{S}}$}mejkal}}, \bibinfo {author} {\bibfnamefont {V.~N.}\ \bibnamefont {Strocov}}, \bibinfo {author} {\bibfnamefont {P.}~\bibnamefont {Constantinou}}, \bibinfo {author} {\bibfnamefont {A.~B.}\ \bibnamefont {Hellenes}}, \bibinfo {author} {\bibfnamefont {R.}~\bibnamefont {Jaeschke~Ubiergo}}, \bibinfo {author} {\bibfnamefont {W.~H.}\ \bibnamefont {Campos}}, \bibinfo {author} {\bibfnamefont {V.~K.}\ \bibnamefont {Bharadwaj}}, \bibinfo {author} {\bibfnamefont {A.}~\bibnamefont {Chakraborty}}, \bibinfo {author} {\bibfnamefont {T.}~\bibnamefont {Denneulin}}, \bibinfo {author} {\bibfnamefont {W.}~\bibnamefont {Shi}}, \bibinfo {author} {\bibfnamefont {R.~E.}\ \bibnamefont {{Dunin-Borkowski}}}, \bibinfo {author} {\bibfnamefont {S.}~\bibnamefont {Das}}, \bibinfo {author} {\bibfnamefont {M.}~\bibnamefont
  {Kl{\"a}ui}}, \bibinfo {author} {\bibfnamefont {J.}~\bibnamefont {Sinova}},\ and\ \bibinfo {author} {\bibfnamefont {M.}~\bibnamefont {Jourdan}},\ }\bibfield  {title} {\bibinfo {title} {Direct observation of altermagnetic band splitting in {{CrSb}} thin films},\ }\href {https://doi.org/10.1038/s41467-024-46476-5} {\bibfield  {journal} {\bibinfo  {journal} {Nat Commun}\ }\textbf {\bibinfo {volume} {15}},\ \bibinfo {pages} {2116} (\bibinfo {year} {2024})}\BibitemShut {NoStop}%
\bibitem [{\citenamefont {Hertz}(1976)}]{hertzQuantumCriticalPhenomena1976}%
  \BibitemOpen
  \bibfield  {author} {\bibinfo {author} {\bibfnamefont {J.~A.}\ \bibnamefont {Hertz}},\ }\bibfield  {title} {\bibinfo {title} {Quantum critical phenomena},\ }\href {https://doi.org/10.1103/PhysRevB.14.1165} {\bibfield  {journal} {\bibinfo  {journal} {Phys. Rev. B}\ }\textbf {\bibinfo {volume} {14}},\ \bibinfo {pages} {1165} (\bibinfo {year} {1976})}\BibitemShut {NoStop}%
\bibitem [{\citenamefont {D{\"u}rrnagel}\ \emph {et~al.}(2024)\citenamefont {D{\"u}rrnagel}, \citenamefont {Hohmann}, \citenamefont {Maity}, \citenamefont {Seufert}, \citenamefont {Klett}, \citenamefont {Klebl},\ and\ \citenamefont {Thomale}}]{durrnagelAltermagneticPhaseTransition2024}%
  \BibitemOpen
  \bibfield  {author} {\bibinfo {author} {\bibfnamefont {M.}~\bibnamefont {D{\"u}rrnagel}}, \bibinfo {author} {\bibfnamefont {H.}~\bibnamefont {Hohmann}}, \bibinfo {author} {\bibfnamefont {A.}~\bibnamefont {Maity}}, \bibinfo {author} {\bibfnamefont {J.}~\bibnamefont {Seufert}}, \bibinfo {author} {\bibfnamefont {M.}~\bibnamefont {Klett}}, \bibinfo {author} {\bibfnamefont {L.}~\bibnamefont {Klebl}},\ and\ \bibinfo {author} {\bibfnamefont {R.}~\bibnamefont {Thomale}},\ }\href {https://doi.org/10.48550/arXiv.2412.14251} {\bibinfo {title} {Altermagnetic phase transition in a {{Lieb}} metal}} (\bibinfo {year} {2024}),\ \Eprint {https://arxiv.org/abs/2412.14251} {arXiv:2412.14251 [cond-mat]} \BibitemShut {NoStop}%
\bibitem [{\citenamefont {He}\ \emph {et~al.}(2025)\citenamefont {He}, \citenamefont {Zhao}, \citenamefont {Luo},\ and\ \citenamefont {Hu}}]{heAltermagnetism$t$$t^prime$$delta$FermiHubbard2025}%
  \BibitemOpen
  \bibfield  {author} {\bibinfo {author} {\bibfnamefont {S.}~\bibnamefont {He}}, \bibinfo {author} {\bibfnamefont {J.}~\bibnamefont {Zhao}}, \bibinfo {author} {\bibfnamefont {H.-G.}\ \bibnamefont {Luo}},\ and\ \bibinfo {author} {\bibfnamefont {S.}~\bibnamefont {Hu}},\ }\href {https://doi.org/10.48550/arXiv.2503.08362} {\bibinfo {title} {Altermagnetism and beyond in the $t$-$t^\prime$-$\delta$ fermi-hubbard model}} (\bibinfo {year} {2025}),\ \Eprint {https://arxiv.org/abs/2503.08362} {arXiv:2503.08362 [cond-mat]} \BibitemShut {NoStop}%
\bibitem [{\citenamefont {Chen}\ \emph {et~al.}(2024)\citenamefont {Chen}, \citenamefont {Duan}, \citenamefont {Liu}, \citenamefont {Cui}, \citenamefont {Yu}, \citenamefont {Xie},\ and\ \citenamefont {Yu}}]{chenSpinExcitationsShastrySutherland2024}%
  \BibitemOpen
  \bibfield  {author} {\bibinfo {author} {\bibfnamefont {H.}~\bibnamefont {Chen}}, \bibinfo {author} {\bibfnamefont {G.}~\bibnamefont {Duan}}, \bibinfo {author} {\bibfnamefont {C.}~\bibnamefont {Liu}}, \bibinfo {author} {\bibfnamefont {Y.}~\bibnamefont {Cui}}, \bibinfo {author} {\bibfnamefont {W.}~\bibnamefont {Yu}}, \bibinfo {author} {\bibfnamefont {Z.~Y.}\ \bibnamefont {Xie}},\ and\ \bibinfo {author} {\bibfnamefont {R.}~\bibnamefont {Yu}},\ }\href {https://doi.org/10.48550/arXiv.2411.00301} {\bibinfo {title} {Spin excitations of the {{Shastry-Sutherland}} model -- altermagnetism and proximate deconfined quantum criticality}} (\bibinfo {year} {2024}),\ \Eprint {https://arxiv.org/abs/2411.00301} {arXiv:2411.00301 [cond-mat]} \BibitemShut {NoStop}%
\bibitem [{\citenamefont {Ferrari}\ and\ \citenamefont {Valent{\'i}}(2024)}]{ferrariAltermagnetismShastrySutherlandLattice2024}%
  \BibitemOpen
  \bibfield  {author} {\bibinfo {author} {\bibfnamefont {F.}~\bibnamefont {Ferrari}}\ and\ \bibinfo {author} {\bibfnamefont {R.}~\bibnamefont {Valent{\'i}}},\ }\bibfield  {title} {\bibinfo {title} {Altermagnetism on the {{Shastry-Sutherland}} lattice},\ }\href {https://doi.org/10.1103/PhysRevB.110.205140} {\bibfield  {journal} {\bibinfo  {journal} {Phys. Rev. B}\ }\textbf {\bibinfo {volume} {110}},\ \bibinfo {pages} {205140} (\bibinfo {year} {2024})}\BibitemShut {NoStop}%
\bibitem [{\citenamefont {Giuli}\ \emph {et~al.}(2025)\citenamefont {Giuli}, \citenamefont {{Mejuto-Zaera}},\ and\ \citenamefont {Capone}}]{giuliAltermagnetismInteractiondrivenItinerant2025}%
  \BibitemOpen
  \bibfield  {author} {\bibinfo {author} {\bibfnamefont {S.}~\bibnamefont {Giuli}}, \bibinfo {author} {\bibfnamefont {C.}~\bibnamefont {{Mejuto-Zaera}}},\ and\ \bibinfo {author} {\bibfnamefont {M.}~\bibnamefont {Capone}},\ }\bibfield  {title} {\bibinfo {title} {Altermagnetism from interaction-driven itinerant magnetism},\ }\href {https://doi.org/10.1103/PhysRevB.111.L020401} {\bibfield  {journal} {\bibinfo  {journal} {Phys. Rev. B}\ }\textbf {\bibinfo {volume} {111}},\ \bibinfo {pages} {L020401} (\bibinfo {year} {2025})}\BibitemShut {NoStop}%
\bibitem [{\citenamefont {Re}(2024)}]{reDiracPointsTopological2024}%
  \BibitemOpen
  \bibfield  {author} {\bibinfo {author} {\bibfnamefont {L.~D.}\ \bibnamefont {Re}},\ }\href {https://doi.org/10.48550/arXiv.2408.14288} {\bibinfo {title} {Dirac points and topological phases in correlated altermagnets}} (\bibinfo {year} {2024}),\ \Eprint {https://arxiv.org/abs/2408.14288} {arXiv:2408.14288 [cond-mat]} \BibitemShut {NoStop}%
\bibitem [{\citenamefont {Kaushal}\ and\ \citenamefont {Franz}(2024)}]{kaushalAltermagnetismModifiedLieb2024}%
  \BibitemOpen
  \bibfield  {author} {\bibinfo {author} {\bibfnamefont {N.}~\bibnamefont {Kaushal}}\ and\ \bibinfo {author} {\bibfnamefont {M.}~\bibnamefont {Franz}},\ }\href {https://doi.org/10.48550/arXiv.2412.16421} {\bibinfo {title} {Altermagnetism in modified {{Lieb}} lattice {{Hubbard}} model}} (\bibinfo {year} {2024}),\ \Eprint {https://arxiv.org/abs/2412.16421} {arXiv:2412.16421 [cond-mat]} \BibitemShut {NoStop}%
\bibitem [{\citenamefont {Yershov}\ \emph {et~al.}(2024)\citenamefont {Yershov}, \citenamefont {Kravchuk}, \citenamefont {Daghofer},\ and\ \citenamefont {{van den Brink}}}]{yershovFluctuationinducedPiezomagnetismLocal2024}%
  \BibitemOpen
  \bibfield  {author} {\bibinfo {author} {\bibfnamefont {K.~V.}\ \bibnamefont {Yershov}}, \bibinfo {author} {\bibfnamefont {V.~P.}\ \bibnamefont {Kravchuk}}, \bibinfo {author} {\bibfnamefont {M.}~\bibnamefont {Daghofer}},\ and\ \bibinfo {author} {\bibfnamefont {J.}~\bibnamefont {{van den Brink}}},\ }\bibfield  {title} {\bibinfo {title} {Fluctuation-induced piezomagnetism in local moment altermagnets},\ }\href {https://doi.org/10.1103/PhysRevB.110.144421} {\bibfield  {journal} {\bibinfo  {journal} {Phys. Rev. B}\ }\textbf {\bibinfo {volume} {110}},\ \bibinfo {pages} {144421} (\bibinfo {year} {2024})}\BibitemShut {NoStop}%
\bibitem [{\citenamefont {Gomonay}\ \emph {et~al.}(2024)\citenamefont {Gomonay}, \citenamefont {Kravchuk}, \citenamefont {{Jaeschke-Ubiergo}}, \citenamefont {Yershov}, \citenamefont {Jungwirth}, \citenamefont {{$\check{\text{S}}$}mejkal}, \citenamefont {van~den Brink},\ and\ \citenamefont {Sinova}}]{gomonayStructureControlDynamics2024}%
  \BibitemOpen
  \bibfield  {author} {\bibinfo {author} {\bibfnamefont {O.}~\bibnamefont {Gomonay}}, \bibinfo {author} {\bibfnamefont {V.~P.}\ \bibnamefont {Kravchuk}}, \bibinfo {author} {\bibfnamefont {R.}~\bibnamefont {{Jaeschke-Ubiergo}}}, \bibinfo {author} {\bibfnamefont {K.~V.}\ \bibnamefont {Yershov}}, \bibinfo {author} {\bibfnamefont {T.}~\bibnamefont {Jungwirth}}, \bibinfo {author} {\bibfnamefont {L.}~\bibnamefont {{$\check{\text{S}}$}mejkal}}, \bibinfo {author} {\bibfnamefont {J.}~\bibnamefont {van~den Brink}},\ and\ \bibinfo {author} {\bibfnamefont {J.}~\bibnamefont {Sinova}},\ }\bibfield  {title} {\bibinfo {title} {Structure, control, and dynamics of altermagnetic textures},\ }\href {https://doi.org/10.1038/s44306-024-00042-3} {\bibfield  {journal} {\bibinfo  {journal} {npj Spintronics}\ }\textbf {\bibinfo {volume} {2}},\ \bibinfo {pages} {1} (\bibinfo {year} {2024})}\BibitemShut {NoStop}%
\bibitem [{\citenamefont {Cônsoli}\ and\ \citenamefont {Vojta}(2025)}]{consoliSUNAltermagnetismLattice2025}%
  \BibitemOpen
  \bibfield  {author} {\bibinfo {author} {\bibfnamefont {P.~M.}\ \bibnamefont {Cônsoli}}\ and\ \bibinfo {author} {\bibfnamefont {M.}~\bibnamefont {Vojta}},\ }\bibfield  {title} {\bibinfo {title} {{$\mathrm{SU}({N})$ Altermagnetism: Lattice Models, Magnon Modes, and Flavor-Split Bands}},\ }\href {https://doi.org/10.1103/PhysRevLett.134.196701} {\bibfield  {journal} {\bibinfo  {journal} {Physical Review Letters}\ }\textbf {\bibinfo {volume} {134}},\ \bibinfo {pages} {196701} (\bibinfo {year} {2025})}\BibitemShut {NoStop}%
\bibitem [{\citenamefont {Canals}(2002)}]{canalsSquareLatticeCheckerboard2002}%
  \BibitemOpen
  \bibfield  {author} {\bibinfo {author} {\bibfnamefont {B.}~\bibnamefont {Canals}},\ }\bibfield  {title} {\bibinfo {title} {{From the square lattice to the checkerboard lattice: Spin-wave and large-$n$ limit analysis}},\ }\href {https://doi.org/10.1103/PhysRevB.65.184408} {\bibfield  {journal} {\bibinfo  {journal} {Phys. Rev. B}\ }\textbf {\bibinfo {volume} {65}},\ \bibinfo {pages} {184408} (\bibinfo {year} {2002})}\BibitemShut {NoStop}%
\bibitem [{\citenamefont {Affleck}(1989)}]{affleckQuantumSpinChains1989}%
  \BibitemOpen
  \bibfield  {author} {\bibinfo {author} {\bibfnamefont {I.}~\bibnamefont {Affleck}},\ }\bibfield  {title} {\bibinfo {title} {Quantum spin chains and the {{Haldane}} gap},\ }\href {https://doi.org/10.1088/0953-8984/1/19/001} {\bibfield  {journal} {\bibinfo  {journal} {J. Phys.: Condens. Matter}\ }\textbf {\bibinfo {volume} {1}},\ \bibinfo {pages} {3047} (\bibinfo {year} {1989})}\BibitemShut {NoStop}%
\bibitem [{\citenamefont {Abanov}(2017)}]{abanovTopologyGeometryQuantum2017}%
  \BibitemOpen
  \bibfield  {author} {\bibinfo {author} {\bibfnamefont {A.}~\bibnamefont {Abanov}},\ }\bibfield  {title} {\bibinfo {title} {Topology, geometry and quantum interference in condensed matter physics},\ }in\ \href {https://doi.org/10.1007/978-981-10-6841-6_12} {\emph {\bibinfo {booktitle} {Topology and {{Condensed Matter Physics}}}}},\ \bibinfo {editor} {edited by\ \bibinfo {editor} {\bibfnamefont {S.~M.}\ \bibnamefont {Bhattacharjee}}, \bibinfo {editor} {\bibfnamefont {M.}~\bibnamefont {Mj}},\ and\ \bibinfo {editor} {\bibfnamefont {A.}~\bibnamefont {Bandyopadhyay}}}\ (\bibinfo  {publisher} {Springer},\ \bibinfo {address} {Singapore},\ \bibinfo {year} {2017})\ pp.\ \bibinfo {pages} {281--331}\BibitemShut {NoStop}%
\bibitem [{\citenamefont {Bhowal}\ and\ \citenamefont {Spaldin}(2024)}]{bhowalFerroicallyOrderedMagnetic2024}%
  \BibitemOpen
  \bibfield  {author} {\bibinfo {author} {\bibfnamefont {S.}~\bibnamefont {Bhowal}}\ and\ \bibinfo {author} {\bibfnamefont {N.~A.}\ \bibnamefont {Spaldin}},\ }\bibfield  {title} {\bibinfo {title} {Ferroically {{Ordered Magnetic Octupoles}} in {$d$}-{{Wave Altermagnets}}},\ }\href {https://doi.org/10.1103/PhysRevX.14.011019} {\bibfield  {journal} {\bibinfo  {journal} {Phys. Rev. X}\ }\textbf {\bibinfo {volume} {14}},\ \bibinfo {pages} {011019} (\bibinfo {year} {2024})}\BibitemShut {NoStop}%
\bibitem [{\citenamefont {McClarty}\ and\ \citenamefont {Rau}(2024)}]{mcclartyLandauTheoryAltermagnetism2024}%
  \BibitemOpen
  \bibfield  {author} {\bibinfo {author} {\bibfnamefont {P.~A.}\ \bibnamefont {McClarty}}\ and\ \bibinfo {author} {\bibfnamefont {J.~G.}\ \bibnamefont {Rau}},\ }\bibfield  {title} {\bibinfo {title} {Landau {{Theory}} of {{Altermagnetism}}},\ }\href {https://doi.org/10.1103/PhysRevLett.132.176702} {\bibfield  {journal} {\bibinfo  {journal} {Phys. Rev. Lett.}\ }\textbf {\bibinfo {volume} {132}},\ \bibinfo {pages} {176702} (\bibinfo {year} {2024})},\ \Eprint {https://arxiv.org/abs/2308.04484} {arXiv:2308.04484 [cond-mat]} \BibitemShut {NoStop}%
\bibitem [{\citenamefont {McClarty}\ \emph {et~al.}(2025)\citenamefont {McClarty}, \citenamefont {Gukasov},\ and\ \citenamefont {Rau}}]{mcclartyObservingAltermagnetismUsing2025}%
  \BibitemOpen
  \bibfield  {author} {\bibinfo {author} {\bibfnamefont {P.~A.}\ \bibnamefont {McClarty}}, \bibinfo {author} {\bibfnamefont {A.}~\bibnamefont {Gukasov}},\ and\ \bibinfo {author} {\bibfnamefont {J.~G.}\ \bibnamefont {Rau}},\ }\bibfield  {title} {\bibinfo {title} {Observing altermagnetism using polarized neutrons},\ }\href {https://doi.org/10.1103/PhysRevB.111.L060405} {\bibfield  {journal} {\bibinfo  {journal} {Phys. Rev. B}\ }\textbf {\bibinfo {volume} {111}},\ \bibinfo {pages} {L060405} (\bibinfo {year} {2025})}\BibitemShut {NoStop}%
\bibitem [{\citenamefont {Witten}(1984)}]{wittenNonabelianBosonizationTwo1984}%
  \BibitemOpen
  \bibfield  {author} {\bibinfo {author} {\bibfnamefont {E.}~\bibnamefont {Witten}},\ }\bibfield  {title} {\bibinfo {title} {Non-abelian bosonization in two dimensions},\ }\href {https://doi.org/10.1007/BF01215276} {\bibfield  {journal} {\bibinfo  {journal} {Commun.Math. Phys.}\ }\textbf {\bibinfo {volume} {92}},\ \bibinfo {pages} {455} (\bibinfo {year} {1984})}\BibitemShut {NoStop}%
\bibitem [{\citenamefont {{Zinn-Justin}}(2021)}]{zinn-justinQuantumFieldTheory2021}%
  \BibitemOpen
  \bibfield  {author} {\bibinfo {author} {\bibfnamefont {J.}~\bibnamefont {{Zinn-Justin}}},\ }\href@noop {} {\emph {\bibinfo {title} {Quantum {{Field Theory}} and {{Critical Phenomena}}: {{Fifth Edition}}}}},\ International {{Series}} of {{Monographs}} on {{Physics}}\ (\bibinfo  {publisher} {Oxford University Press},\ \bibinfo {year} {2021})\BibitemShut {NoStop}%
\bibitem [{\citenamefont {Nelson}\ and\ \citenamefont {Pelcovits}(1977)}]{nelsonMomentumshellRecursionRelations1977}%
  \BibitemOpen
  \bibfield  {author} {\bibinfo {author} {\bibfnamefont {D.~R.}\ \bibnamefont {Nelson}}\ and\ \bibinfo {author} {\bibfnamefont {R.~A.}\ \bibnamefont {Pelcovits}},\ }\bibfield  {title} {\bibinfo {title} {Momentum-shell recursion relations, anisotropic spins, and liquid crystals in $2+\ensuremath{\epsilon}$ dimensions},\ }\href {https://doi.org/10.1103/PhysRevB.16.2191} {\bibfield  {journal} {\bibinfo  {journal} {Phys. Rev. B}\ }\textbf {\bibinfo {volume} {16}},\ \bibinfo {pages} {2191} (\bibinfo {year} {1977})}\BibitemShut {NoStop}%
\bibitem [{\citenamefont {Sachdev}(2011)}]{sachdevQuantumPhaseTransitions2011}%
  \BibitemOpen
  \bibfield  {author} {\bibinfo {author} {\bibfnamefont {S.}~\bibnamefont {Sachdev}},\ }\href {https://doi.org/10.1017/CBO9780511973765} {\emph {\bibinfo {title} {Quantum {{Phase Transitions}}}}},\ \bibinfo {edition} {2nd}\ ed.\ (\bibinfo  {publisher} {Cambridge University Press},\ \bibinfo {address} {Cambridge},\ \bibinfo {year} {2011})\BibitemShut {NoStop}%
\bibitem [{\citenamefont {Das}\ \emph {et~al.}(2024)\citenamefont {Das}, \citenamefont {Leeb}, \citenamefont {Knolle},\ and\ \citenamefont {Knap}}]{dasRealizingAltermagnetismFermiHubbard2024}%
  \BibitemOpen
  \bibfield  {author} {\bibinfo {author} {\bibfnamefont {P.}~\bibnamefont {Das}}, \bibinfo {author} {\bibfnamefont {V.}~\bibnamefont {Leeb}}, \bibinfo {author} {\bibfnamefont {J.}~\bibnamefont {Knolle}},\ and\ \bibinfo {author} {\bibfnamefont {M.}~\bibnamefont {Knap}},\ }\bibfield  {title} {\bibinfo {title} {Realizing {{Altermagnetism}} in {{Fermi-Hubbard Models}} with {{Ultracold Atoms}}},\ }\href {https://doi.org/10.1103/PhysRevLett.132.263402} {\bibfield  {journal} {\bibinfo  {journal} {Phys. Rev. Lett.}\ }\textbf {\bibinfo {volume} {132}},\ \bibinfo {pages} {263402} (\bibinfo {year} {2024})}\BibitemShut {NoStop}%
\bibitem [{\citenamefont {Bose}\ \emph {et~al.}(2024)\citenamefont {Bose}, \citenamefont {Vadnais},\ and\ \citenamefont {Paramekanti}}]{boseAltermagnetismSuperconductivityMultiorbital2024}%
  \BibitemOpen
  \bibfield  {author} {\bibinfo {author} {\bibfnamefont {A.}~\bibnamefont {Bose}}, \bibinfo {author} {\bibfnamefont {S.}~\bibnamefont {Vadnais}},\ and\ \bibinfo {author} {\bibfnamefont {A.}~\bibnamefont {Paramekanti}},\ }\bibfield  {title} {\bibinfo {title} {{Altermagnetism and superconductivity in a multiorbital $t\ensuremath{-}J$ model}},\ }\href {https://doi.org/10.1103/PhysRevB.110.205120} {\bibfield  {journal} {\bibinfo  {journal} {Phys. Rev. B}\ }\textbf {\bibinfo {volume} {110}},\ \bibinfo {pages} {205120} (\bibinfo {year} {2024})}\BibitemShut {NoStop}%
\bibitem [{\citenamefont {{Meng-Han}}\ \emph {et~al.}(2024)\citenamefont {{Meng-Han}}, \citenamefont {Xuan},\ and\ \citenamefont {Yao}}]{meng-hanDiracPointsWeyl2024}%
  \BibitemOpen
  \bibfield  {author} {\bibinfo {author} {\bibfnamefont {Z.}~\bibnamefont {{Meng-Han}}}, \bibinfo {author} {\bibfnamefont {G.}~\bibnamefont {Xuan}},\ and\ \bibinfo {author} {\bibfnamefont {D.-X.}\ \bibnamefont {Yao}},\ }\href {https://doi.org/10.48550/arXiv.2412.03657} {\bibinfo {title} {Dirac points and {{Weyl}} phase in a honeycomb altermagnet}} (\bibinfo {year} {2024}),\ \Eprint {https://arxiv.org/abs/2412.03657} {arXiv:2412.03657 [cond-mat]} \BibitemShut {NoStop}%
\bibitem [{\citenamefont {Antonenko}\ \emph {et~al.}(2025)\citenamefont {Antonenko}, \citenamefont {Fernandes},\ and\ \citenamefont {Venderbos}}]{antonenkoMirrorChernBands2025}%
  \BibitemOpen
  \bibfield  {author} {\bibinfo {author} {\bibfnamefont {D.~S.}\ \bibnamefont {Antonenko}}, \bibinfo {author} {\bibfnamefont {R.~M.}\ \bibnamefont {Fernandes}},\ and\ \bibinfo {author} {\bibfnamefont {J.~W.~F.}\ \bibnamefont {Venderbos}},\ }\bibfield  {title} {\bibinfo {title} {Mirror {{Chern Bands}} and {{Weyl Nodal Loops}} in {{Altermagnets}}},\ }\href {https://doi.org/10.1103/PhysRevLett.134.096703} {\bibfield  {journal} {\bibinfo  {journal} {Physical Review Letters}\ }\textbf {\bibinfo {volume} {134}},\ \bibinfo {pages} {096703} (\bibinfo {year} {2025})}\BibitemShut {NoStop}%
\bibitem [{\citenamefont {Abanov}\ \emph {et~al.}(2003)\citenamefont {Abanov}, \citenamefont {Chubukov},\ and\ \citenamefont {Schmalian}}]{abanovQuantumcriticalTheorySpinfermion2003}%
  \BibitemOpen
  \bibfield  {author} {\bibinfo {author} {\bibfnamefont {{\relax Ar}.}~\bibnamefont {Abanov}}, \bibinfo {author} {\bibfnamefont {A.~V.}\ \bibnamefont {Chubukov}},\ and\ \bibinfo {author} {\bibfnamefont {J.}~\bibnamefont {Schmalian}},\ }\bibfield  {title} {\bibinfo {title} {Quantum-critical theory of the spin-fermion model and its application to cuprates: {{Normal}} state analysis},\ }\href {https://doi.org/10.1080/0001873021000057123} {\bibfield  {journal} {\bibinfo  {journal} {Advances in Physics}\ }\textbf {\bibinfo {volume} {52}},\ \bibinfo {pages} {119} (\bibinfo {year} {2003})}\BibitemShut {NoStop}%
\bibitem [{\citenamefont {Metlitski}\ and\ \citenamefont {Sachdev}(2010)}]{metlitskiQuantumPhaseTransitions2010}%
  \BibitemOpen
  \bibfield  {author} {\bibinfo {author} {\bibfnamefont {M.~A.}\ \bibnamefont {Metlitski}}\ and\ \bibinfo {author} {\bibfnamefont {S.}~\bibnamefont {Sachdev}},\ }\bibfield  {title} {\bibinfo {title} {Quantum phase transitions of metals in two spatial dimensions. {{II}}. {{Spin}} density wave order},\ }\href {https://doi.org/10.1103/PhysRevB.82.075128} {\bibfield  {journal} {\bibinfo  {journal} {Phys. Rev. B}\ }\textbf {\bibinfo {volume} {82}},\ \bibinfo {pages} {075128} (\bibinfo {year} {2010})}\BibitemShut {NoStop}%
\bibitem [{\citenamefont {Lee}(2018)}]{leeRecentDevelopmentsNonFermi2018}%
  \BibitemOpen
  \bibfield  {author} {\bibinfo {author} {\bibfnamefont {S.-S.}\ \bibnamefont {Lee}},\ }\bibfield  {title} {\bibinfo {title} {Recent {{Developments}} in {{Non-Fermi Liquid Theory}}},\ }\href {https://doi.org/10.1146/annurev-conmatphys-031016-025531} {\bibfield  {journal} {\bibinfo  {journal} {Annual Review of Condensed Matter Physics}\ }\textbf {\bibinfo {volume} {9}},\ \bibinfo {pages} {227} (\bibinfo {year} {2018})}\BibitemShut {NoStop}%
\bibitem [{\citenamefont {Herbut}(2001)}]{herbutQuantumCriticalPoints2001}%
  \BibitemOpen
  \bibfield  {author} {\bibinfo {author} {\bibfnamefont {I.~F.}\ \bibnamefont {Herbut}},\ }\bibfield  {title} {\bibinfo {title} {Quantum critical points with the coulomb interaction and the dynamical exponent: When and why $\mathit{z}\phantom{\rule{0ex}{0ex}}=\phantom{\rule{0ex}{0ex}}1$},\ }\href {https://doi.org/10.1103/PhysRevLett.87.137004} {\bibfield  {journal} {\bibinfo  {journal} {Phys. Rev. Lett.}\ }\textbf {\bibinfo {volume} {87}},\ \bibinfo {pages} {137004} (\bibinfo {year} {2001})}\BibitemShut {NoStop}%
\bibitem [{\citenamefont {Herbut}(2006)}]{herbutInteractionsPhaseTransitions2006}%
  \BibitemOpen
  \bibfield  {author} {\bibinfo {author} {\bibfnamefont {I.~F.}\ \bibnamefont {Herbut}},\ }\bibfield  {title} {\bibinfo {title} {Interactions and {{Phase Transitions}} on {{Graphene}}'s {{Honeycomb Lattice}}},\ }\href {https://doi.org/10.1103/PhysRevLett.97.146401} {\bibfield  {journal} {\bibinfo  {journal} {Phys. Rev. Lett.}\ }\textbf {\bibinfo {volume} {97}},\ \bibinfo {pages} {146401} (\bibinfo {year} {2006})}\BibitemShut {NoStop}%
\bibitem [{\citenamefont {Son}(2007)}]{sonQuantumCriticalPoint2007}%
  \BibitemOpen
  \bibfield  {author} {\bibinfo {author} {\bibfnamefont {D.~T.}\ \bibnamefont {Son}},\ }\bibfield  {title} {\bibinfo {title} {Quantum critical point in graphene approached in the limit of infinitely strong {{Coulomb}} interaction},\ }\href {https://doi.org/10.1103/PhysRevB.75.235423} {\bibfield  {journal} {\bibinfo  {journal} {Phys. Rev. B}\ }\textbf {\bibinfo {volume} {75}},\ \bibinfo {pages} {235423} (\bibinfo {year} {2007})}\BibitemShut {NoStop}%
\bibitem [{\citenamefont {Hsiao}\ and\ \citenamefont {Son}(2017)}]{hsiaoDualityUniversalTransport2017}%
  \BibitemOpen
  \bibfield  {author} {\bibinfo {author} {\bibfnamefont {W.-H.}\ \bibnamefont {Hsiao}}\ and\ \bibinfo {author} {\bibfnamefont {D.~T.}\ \bibnamefont {Son}},\ }\bibfield  {title} {\bibinfo {title} {Duality and universal transport in mixed-dimension electrodynamics},\ }\href {https://doi.org/10.1103/PhysRevB.96.075127} {\bibfield  {journal} {\bibinfo  {journal} {Physical Review B}\ }\textbf {\bibinfo {volume} {96}},\ \bibinfo {pages} {075127} (\bibinfo {year} {2017})}\BibitemShut {NoStop}%
\bibitem [{\citenamefont {Zhang}\ \emph {et~al.}(2024)\citenamefont {Zhang}, \citenamefont {Cheng}, \citenamefont {Yin}, \citenamefont {Liu}, \citenamefont {Deng}, \citenamefont {Qiao}, \citenamefont {Shi}, \citenamefont {Zhang}, \citenamefont {Lin}, \citenamefont {Liu}, \citenamefont {Ye}, \citenamefont {Huang}, \citenamefont {Meng}, \citenamefont {Zhang}, \citenamefont {Okuda}, \citenamefont {Shimada}, \citenamefont {Cui}, \citenamefont {Zhao}, \citenamefont {Cao}, \citenamefont {Qiao}, \citenamefont {Liu},\ and\ \citenamefont {Chen}}]{zhangCrystalsymmetrypairedSpinvalleyLocking2024}%
  \BibitemOpen
  \bibfield  {author} {\bibinfo {author} {\bibfnamefont {F.}~\bibnamefont {Zhang}}, \bibinfo {author} {\bibfnamefont {X.}~\bibnamefont {Cheng}}, \bibinfo {author} {\bibfnamefont {Z.}~\bibnamefont {Yin}}, \bibinfo {author} {\bibfnamefont {C.}~\bibnamefont {Liu}}, \bibinfo {author} {\bibfnamefont {L.}~\bibnamefont {Deng}}, \bibinfo {author} {\bibfnamefont {Y.}~\bibnamefont {Qiao}}, \bibinfo {author} {\bibfnamefont {Z.}~\bibnamefont {Shi}}, \bibinfo {author} {\bibfnamefont {S.}~\bibnamefont {Zhang}}, \bibinfo {author} {\bibfnamefont {J.}~\bibnamefont {Lin}}, \bibinfo {author} {\bibfnamefont {Z.}~\bibnamefont {Liu}}, \bibinfo {author} {\bibfnamefont {M.}~\bibnamefont {Ye}}, \bibinfo {author} {\bibfnamefont {Y.}~\bibnamefont {Huang}}, \bibinfo {author} {\bibfnamefont {X.}~\bibnamefont {Meng}}, \bibinfo {author} {\bibfnamefont {C.}~\bibnamefont {Zhang}}, \bibinfo {author} {\bibfnamefont {T.}~\bibnamefont {Okuda}}, \bibinfo {author} {\bibfnamefont {K.}~\bibnamefont {Shimada}}, \bibinfo {author} {\bibfnamefont
  {S.}~\bibnamefont {Cui}}, \bibinfo {author} {\bibfnamefont {Y.}~\bibnamefont {Zhao}}, \bibinfo {author} {\bibfnamefont {G.-H.}\ \bibnamefont {Cao}}, \bibinfo {author} {\bibfnamefont {S.}~\bibnamefont {Qiao}}, \bibinfo {author} {\bibfnamefont {J.}~\bibnamefont {Liu}},\ and\ \bibinfo {author} {\bibfnamefont {C.}~\bibnamefont {Chen}},\ }\href {https://doi.org/10.48550/arXiv.2407.19555} {\bibinfo {title} {Crystal-symmetry-paired spin-valley locking in a layered room-temperature antiferromagnet}} (\bibinfo {year} {2024}),\ \Eprint {https://arxiv.org/abs/2407.19555} {arXiv:2407.19555 [cond-mat]} \BibitemShut {NoStop}%
\bibitem [{\citenamefont {Bauer}\ \emph {et~al.}(2015)\citenamefont {Bauer}, \citenamefont {R{\"u}ckriegel}, \citenamefont {Sharma},\ and\ \citenamefont {Kopietz}}]{bauerNonperturbativeRenormalizationGroup2015}%
  \BibitemOpen
  \bibfield  {author} {\bibinfo {author} {\bibfnamefont {C.}~\bibnamefont {Bauer}}, \bibinfo {author} {\bibfnamefont {A.}~\bibnamefont {R{\"u}ckriegel}}, \bibinfo {author} {\bibfnamefont {A.}~\bibnamefont {Sharma}},\ and\ \bibinfo {author} {\bibfnamefont {P.}~\bibnamefont {Kopietz}},\ }\bibfield  {title} {\bibinfo {title} {Nonperturbative renormalization group calculation of quasiparticle velocity and dielectric function of graphene},\ }\href {https://doi.org/10.1103/PhysRevB.92.121409} {\bibfield  {journal} {\bibinfo  {journal} {Phys. Rev. B}\ }\textbf {\bibinfo {volume} {92}},\ \bibinfo {pages} {121409} (\bibinfo {year} {2015})}\BibitemShut {NoStop}%
\bibitem [{\citenamefont {Lundemo}\ \emph {et~al.}(2025)\citenamefont {Lundemo}, \citenamefont {Nogueira},\ and\ \citenamefont {Sudb{\o}}}]{lundemoTopologydrivenDeconfinedQuantum2025}%
  \BibitemOpen
  \bibfield  {author} {\bibinfo {author} {\bibfnamefont {S.~D.}\ \bibnamefont {Lundemo}}, \bibinfo {author} {\bibfnamefont {F.~S.}\ \bibnamefont {Nogueira}},\ and\ \bibinfo {author} {\bibfnamefont {A.}~\bibnamefont {Sudb{\o}}},\ }\bibfield  {title} {\bibinfo {title} {Topology-driven deconfined quantum criticality in magnetic bilayers},\ }\href {https://doi.org/10.1103/PhysRevB.111.075158} {\bibfield  {journal} {\bibinfo  {journal} {Phys. Rev. B}\ }\textbf {\bibinfo {volume} {111}},\ \bibinfo {pages} {075158} (\bibinfo {year} {2025})}\BibitemShut {NoStop}%
\bibitem [{\citenamefont {Appelquist}\ \emph {et~al.}(1988)\citenamefont {Appelquist}, \citenamefont {Nash},\ and\ \citenamefont {Wijewardhana}}]{appelquistCriticalBehavior2+1Dimensional1988}%
  \BibitemOpen
  \bibfield  {author} {\bibinfo {author} {\bibfnamefont {T.}~\bibnamefont {Appelquist}}, \bibinfo {author} {\bibfnamefont {D.}~\bibnamefont {Nash}},\ and\ \bibinfo {author} {\bibfnamefont {L.~C.~R.}\ \bibnamefont {Wijewardhana}},\ }\bibfield  {title} {\bibinfo {title} {Critical {{Behavior}} in (2+1)-{{Dimensional QED}}},\ }\href {https://doi.org/10.1103/PhysRevLett.60.2575} {\bibfield  {journal} {\bibinfo  {journal} {Phys. Rev. Lett.}\ }\textbf {\bibinfo {volume} {60}},\ \bibinfo {pages} {2575} (\bibinfo {year} {1988})}\BibitemShut {NoStop}%
\bibitem [{\citenamefont {Appelquist}\ \emph {et~al.}(1986)\citenamefont {Appelquist}, \citenamefont {Bowick}, \citenamefont {Karabali},\ and\ \citenamefont {Wijewardhana}}]{appelquistSpontaneousChiralsymmetryBreaking1986}%
  \BibitemOpen
  \bibfield  {author} {\bibinfo {author} {\bibfnamefont {T.~W.}\ \bibnamefont {Appelquist}}, \bibinfo {author} {\bibfnamefont {M.}~\bibnamefont {Bowick}}, \bibinfo {author} {\bibfnamefont {D.}~\bibnamefont {Karabali}},\ and\ \bibinfo {author} {\bibfnamefont {L.~C.~R.}\ \bibnamefont {Wijewardhana}},\ }\bibfield  {title} {\bibinfo {title} {Spontaneous chiral-symmetry breaking in three-dimensional {{QED}}},\ }\href {https://doi.org/10.1103/PhysRevD.33.3704} {\bibfield  {journal} {\bibinfo  {journal} {Phys. Rev. D}\ }\textbf {\bibinfo {volume} {33}},\ \bibinfo {pages} {3704} (\bibinfo {year} {1986})}\BibitemShut {NoStop}%
\end{thebibliography}%
	
\end{document}